\documentclass[twocolumn,amsmath,amssymb,amsfonts,superscriptaddress,floatfix,showpacs,prd,aps]{revtex4}
\newcommand{\del}{\partial}
\newcommand{\Tr}{\mathrm{Tr}}

\newcommand{\vev}[1]{\langle{#1}\rangle}

\usepackage{graphicx,color,mathrsfs,flafter
}

\begin{document}
\title{\bf The renormalization group and quark number fluctuations in
  the Polyakov loop extended quark-meson model at finite baryon
  density}

\author{V. Skokov} \email[E-Mail:]{V.Skokov@gsi.de} \affiliation{%
  GSI Helmholtzzentrum f\"ur Schwerionenforschung, D-64291 Darmstadt,
  Germany}
\author{B.~Friman} \affiliation{%
  GSI Helmholtzzentrum f\"ur Schwerionenforschung, D-64291 Darmstadt,
  Germany}
\author{K.~Redlich} \affiliation{%
  Institute of Theoretical Physics, University of Wroclaw, PL--50204
  Wroc\l aw, Poland} \affiliation{%
  Theory Division, CERN, CH-1211 Geneva 23, Switzerland}



\begin{abstract}
  Thermodynamics and the phase structure of the Polyakov loop-extended
  two flavors chiral quark--meson (PQM) model is explored beyond the
  mean-field approximation.  The analysis of the PQM model is based on
  the functional renormalization group (FRG) method.  We formulate and
  solve the renormalization group flow equation for the
  scale-dependent thermodynamic potential in the presence of the
  gluonic background field at finite temperature and density.  We
  determine the phase diagram of the PQM model in the FRG approach and
  discuss its modification in comparison with the one obtained under 
	the mean-field approximation.  We
  focus on  properties of the net-quark number density fluctuations
  as well as  their higher moments and discuss the influence of 
  non-perturbative effects
  on their properties near the chiral crossover transition. We show,
  that with an increasing net-quark number density the higher order
  moments  exhibit a peculiar structure near the phase transition.
  We also consider ratios of different moments of the
  net-quark number density and discuss their role as probes of 
  deconfinement and chiral phase transitions.

\end{abstract}

\maketitle

\section{Introduction}
Thermodynamic properties of a strongly interacting matter at nonzero
baryon density and high temperature were quantified numerically within
Lattice QCD (LQCD) \cite{L_lgt1,L_lgt2,L_lgt3}.  The LQCD results
demonstrate that QCD exhibits both dynamical chiral symmetry breaking
and confinement at finite temperature and densities.  The LQCD
equation of state indicates a clear separation between the confined
hadronic phase and deconfined phase of quark--gluon plasma.  However,
since the thermodynamics at large baryon densities and near the chiral
limit is still not accessible from the first principle LQCD
calculations, many phenomenological models and effective theories have
been developed
~\cite{Gocksch,Buballa:review,Meisinger,Fukushima,Mocsy,PNJL,CS,DLS,Megias,IK,Fukushima:strong,Schaefer:PQM,kap,sasaki,model}.

The hadronic properties at low energy as well as the nature of the
chiral phase transition at finite temperature and densities have been
successfully explored and described in such effective models.
The physics of color deconfinement and its relation to the chiral
symmetry breaking has been recently studied in terms of effective
models.  The idea to extend the existing chiral Lagrangians such as
the Nambu--Jona--Lasinio or the quark--meson, by introducing coupling
of quarks to uniform temporal background gauge fields (Polyakov loops)
was an important step forward in these studies
~\cite{Fukushima,Schaefer:PQM}.

It was shown that the Polyakov loop extended Nambu--Jona--Lasinio
(PNJL) \cite{PNJL} or quark--meson (PQM) \cite{Schaefer:PQM} models
can reproduce essential properties of the QCD thermodynamics obtained
in the LQCD already within the mean-field approximation. However,
to correctly account for the critical behavior and scaling properties
near the chiral phase transition one needs to go beyond the
mean-field approximation and include quantum fluctuations and
non-perturbative dynamics. This can be achieved by using the methods
based on the functional renormalization group
(FRG)~\cite{Wetterich,Morris,Ellwanger,Berges:review,Schaefer:2006ds,SFR,Herbst:2010rf}.

Following our previous work~\cite{Skokov:2010wb} we use a  truncation
of the PQM model which is suitable for the functional renormalization
group analysis to describe the thermodynamics beyond the mean-field
approximation. The functional renormalization group approach in the PQM
model is used to take into account fluctuations of the meson fields,
while the Polyakov loop is treated as a background field on the
mean-field level.
In contrast to the previous work~\cite{Skokov:2010wb} we extend our
calculations to the finite chemical potential.  We determine the phase
diagram and the position of the critical end point (CEP) in the PQM
model by exploring the dependencies of the chiral order parameter and
the quark number susceptibility on thermal variables.  We calculate
the moments (cumulants) of the net-quark number density fluctuations ($c_n$)
at finite temperature and chemical potential in the presence
of mesonic fluctuations. 
We discuss the
influence of non-perturbative effects on properties of the first four
$c_n$-moments near the chiral crossover transition. We show that
$c_n$-cumulants exhibit a peculiar  structure and for sufficiently large
values of the chemical potential can be negative
near to  the crossover   transition.  
We calculate
the ratios $c_3/c_1$ and $c_4/c_2$ and discuss their roles as probes of
  deconfinement and chiral phase transitions.
We summarized properties of different susceptibilities near the chiral
phase transition at finite net-quark density within the Landau { mean-field}
and the scaling theories.

\section{The Polyakov-quark-meson model}\label{sec:pqm}

The model which { is used in this paper}  to explore the chiral phase
transition at finite temperature and density within the FRG approach is
the Polyakov loop-extended two flavor quark--meson model.  In general,
the PQM model, being an effective realisation of the low--energy
sector of the QCD, cannot describe confinement phenomena because the
local $SU(N_c)$ invariance of the QCD is replaced by the global
symmetry.  However, it was argued that by connecting the chiral
quark--meson (QM) model with the Polyakov loop potential the confining
properties of QCD can be approximately accounted for~\cite{Fukushima,
  Fukushima:strong, Schaefer:PQM}.

The Lagrangian of the PQM model reads \cite{Schaefer:PQM}
\begin{eqnarray}\label{eq:pqm_lagrangian}
  {\cal L} &=& \bar{q} \, \left[i\gamma^\mu {D}_\mu  - g (\sigma + i \gamma_5
    \vec \tau \vec \pi )\right]\,q
  +\frac 1 2 (\partial_\mu \sigma)^2+ \frac{ 1}{2}
  (\partial_\mu \vec \pi)^2
  \nonumber \\
  && \qquad - U(\sigma, \vec \pi )  -{\cal U}(\ell,\ell^{*})\ .
\end{eqnarray}
The coupling between the effective gluon field and quarks is
implemented through the covariant derivative
\begin{equation}
  D_{\mu}=\del_{\mu}-iA_{\mu},
\end{equation}
where $A_\mu=g\,A_\mu^a\,\lambda^a/2$. The spatial components of the
gluon field are neglected, i.e. $A_{\mu}=\delta_{\mu0}A_0$.  Moreover,
${\cal U}(\ell,\ell^{*})$ is the effective potential for the gluon
field expressed in terms of the thermal expectation values of the
color trace of the Polyakov loop and its conjugate
\begin{equation}
  \ell=\frac{1}{N_c}\vev{\Tr_c L(\vec{x})},\quad \ell^{*}=\frac{1}{N_c}\vev{\Tr_c
    L^{\dagger}(\vec{x})},
\end{equation}
with
\begin{eqnarray}
  L(\vec x)={\mathcal P} \exp \left[ i \int_0^\beta d\tau A_4(\vec x , \tau)
  \right]\,,
\end{eqnarray}
where ${\mathcal P}$ stands for the path ordering, $\beta=1/T$ and
$A_4=i\,A_0$.  In the $O(4)$ representation the meson field is
introduced as $\phi=(\sigma,\vec{\pi})$ and  the corresponding
$SU(2)_L\otimes SU(2)_R$ chiral representation is defined by
$\sigma+i\vec{\tau}\cdot\vec{\pi}\gamma_5$.

The purely mesonic potential of the model, $U(\sigma,\vec{\pi})$, is
defined as
\begin{equation}
  U(\sigma,\vec{\pi})=\frac{\lambda}{4}\left(\sigma^2+\vec{\pi}
    ^2-v^2\right)^2-c\sigma,
\end{equation}
while the effective potential of the gluon field is parametrized in
such a way as to preserve the $Z(3)$ invariance:
\begin{equation}
  \frac{{\cal U}(\ell,\ell^{*})}{T^4}=
  -\frac{b_2(T)}{2}\ell^{*}\ell
  -\frac{b_3}{6}(\ell^3 + \ell^{*3})
  +\frac{b_4}{4}(\ell^{*}\ell)^2\,\label{eff_potential}.
\end{equation}
The parameters,
\begin{eqnarray}
  \hspace{-4ex}
  b_2(T) &=& a_0  + a_1 \left(\frac{T_0}{T}\right) + a_2
  \left(\frac{T_0}{T}\right)^2 + a_3 \left(\frac{T_0}{T}\right)^3\,
\end{eqnarray}
with $a_0 = 6.75$, $a_1 = -1.95$, $a_2 = 2.625$, $a_3 = -7.44$, $b_3 =
0.75$ and $b_4 = 7.5$ were chosen to reproduce the equation of state
of the pure SU$_c$(3) lattice gauge theory.  Consequently, at
$T_0\simeq 270$ MeV the potential~(\ref{eff_potential}) yields the
first-order deconfinement phase transition.  Several alternative
parametrizations of the effective potential of the gluon field were
also explored {in Refs.}~\cite{Ratti:2007jf,Fukushima:2008wg,Schaefer:2009ui}.

\subsection{The FRG method in the PQM model}\label{sec:rg}

In order to formulate a non-perturbative thermodynamics within the PQM
model we adopt a method based on the functional renormalization group (FRG).
The FRG is based on an infrared regularization with the momentum scale
parameter of the full propagator which turns the corresponding
effective action into the scale $k$-dependent functional
$\Gamma_k$~\cite{Wetterich, Morris, Ellwanger, Berges:review}.

In the PQM model the formulation of the FRG flow
equation 
would require  implementation of the Polyakov loop as a dynamical
field. However, in the current calculation we treat the Polyakov loop
as a background field which is introduced selfconsistently on the
mean-field level.
\begin{figure*}
  \includegraphics*[width=7.5cm]{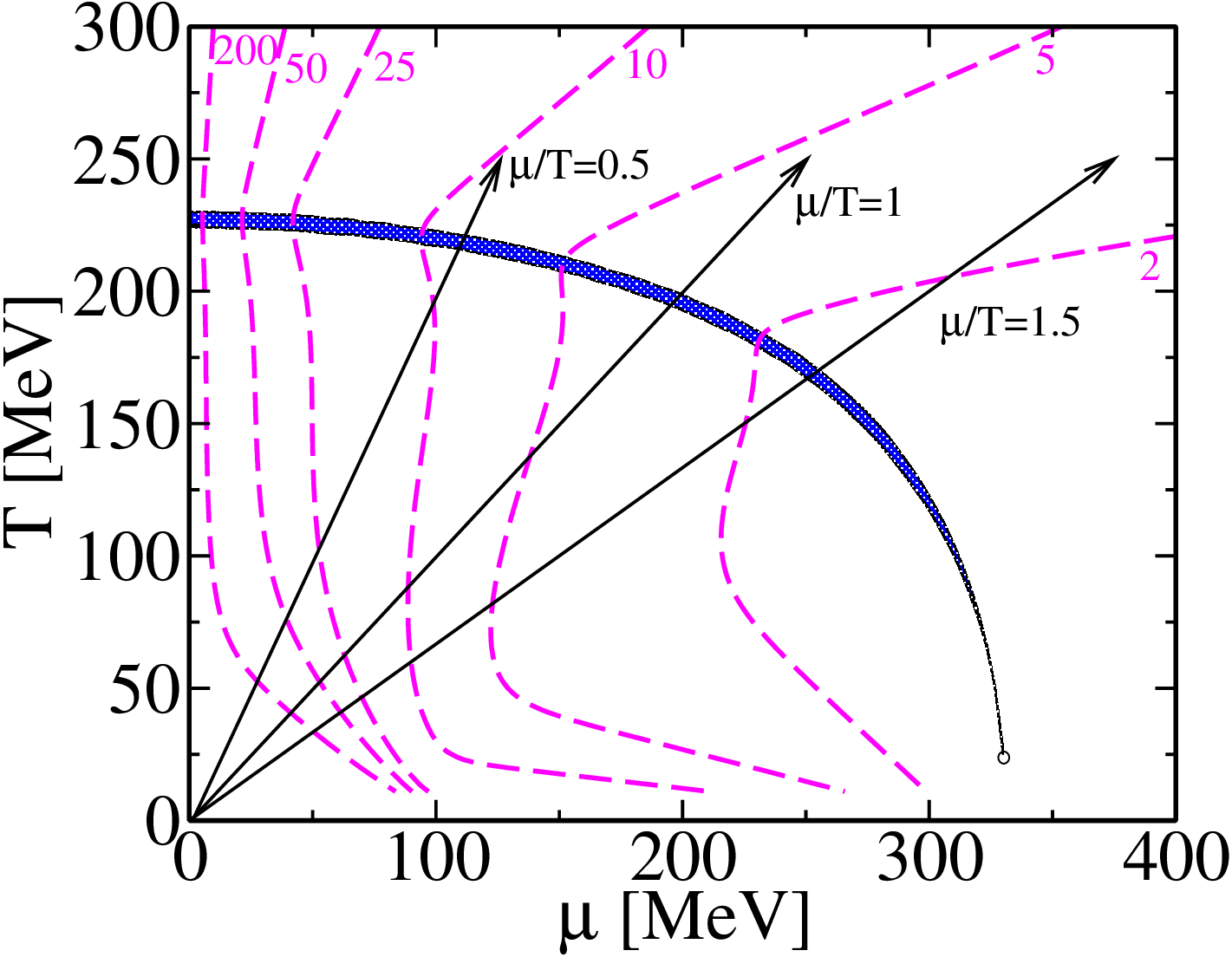}
  \includegraphics*[width=7.5cm]{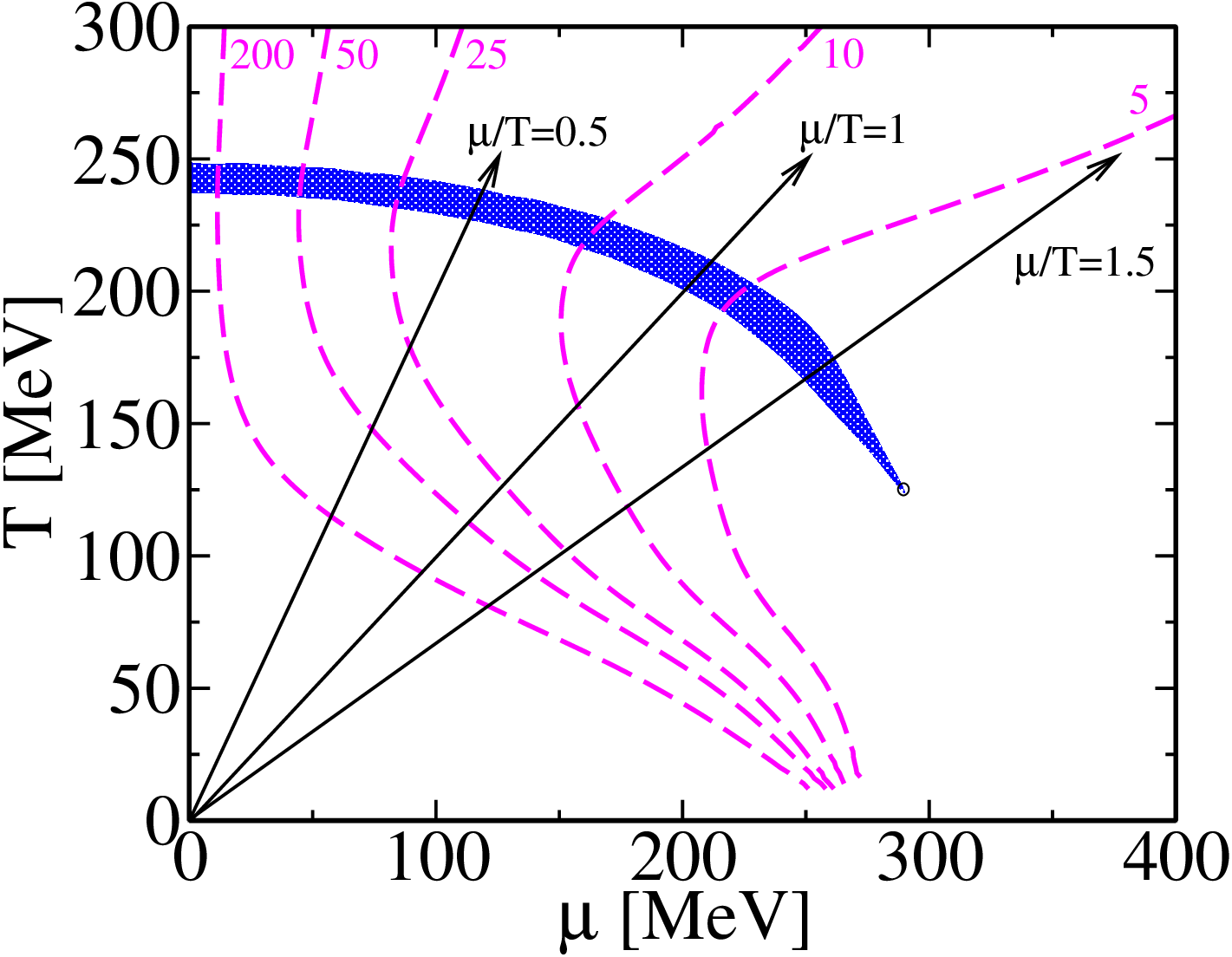}
  \caption { The phase diagrams for the PQM model in the mean-field
    approximation (left panel) and in the FRG approach (right
    panel). The shaded regions are defined by $5\%$-deviations of the
    temperature derivative of the chiral order parameter,
    $d\sigma/dT$, from its maximal value.  The arrows show the lines
    corresponding to different values of $\mu/T$. The dashed curves
    indicate isentropes for $s/n_{q}=$ 2, 5, 10, 25, 50, 200.  }
  \label{fig:PD}
\end{figure*}
\begin{widetext}
  Following our previous work~\cite{Skokov:2010wb} we formulate the
  flow equation for the scale-dependent grand canonical potential for
  the quark and mesonic subsystems
  \begin{eqnarray}\label{eq:frg_flow}
    \del_k \Omega_k(\ell, \ell^*; T,\mu)&=&\frac{k^4}{12\pi^2}
    \left\{ \frac{3}{E_\pi} \Bigg[ 1+2n_B(E_\pi;T)\Bigg]
      +\frac{1}{E_\sigma} \Bigg[ 1+2n_B(E_\sigma;T)
      \Bigg]   \right. \\ \nonumber && \left. -\frac{4 N_c N_f}{E_q} \Bigg[ 1-
      N(\ell,\ell^*;T,\mu)-\bar{N}(\ell,\ell^*;T,\mu)\Bigg] \right\}.
  \end{eqnarray}
  Here $n_B(E_{\pi,\sigma};T)$ is the bosonic distribution function
  \begin{equation*}
    n_B(E_{\pi,\sigma};T)=\frac{1}{\exp({E_{\pi,\sigma}/T})-1}
  \end{equation*}
  with the pion and sigma energies given by 
  \begin{equation*}
    E_\pi = \sqrt{k^2+\overline{\Omega}^{\,\prime}_k}\;~,~ E_\sigma
    =\sqrt{k^2+\overline{\Omega}^{\,\prime}_k+2\rho\,\overline{\Omega}^{\,
        \prime\prime} _k};
  \end{equation*}
  where the primes denote derivatives with respect to $\rho$ and
  $\overline{\Omega}=\Omega+c\sigma$.
  The functions $N(\ell,\ell^*;T,\mu)$ and
  $\bar{N}(\ell,\ell^*;T,\mu)$ defined by
  \begin{eqnarray}\label{n1}
    N(\ell,\ell^*;T,\mu)&=&\frac{1+2\ell^*\exp[\beta(E_q-\mu)]+\ell \exp[2\beta(E_q-\mu)]}{1+3\ell \exp[2\beta(E_q-\mu)]+
      3\ell^*\exp[\beta(E_q-\mu)]+\exp[3\beta(E_q-\mu)]},  \\
    \bar{N}(\ell,\ell^*;T,\mu)&=&N(\ell^*,\ell;T,-\mu)
    \label{n2}
  \end{eqnarray}
  are fermionic distributions 
  modified because of the  coupling to gluons. {The quark energy is defined by}
  \begin{equation}
    \label{dispertion}
    E_q =\sqrt{k^2+2g^2\rho}.
  \end{equation}
\end{widetext}

\begin{figure*}
  \includegraphics*[width=7.5cm]{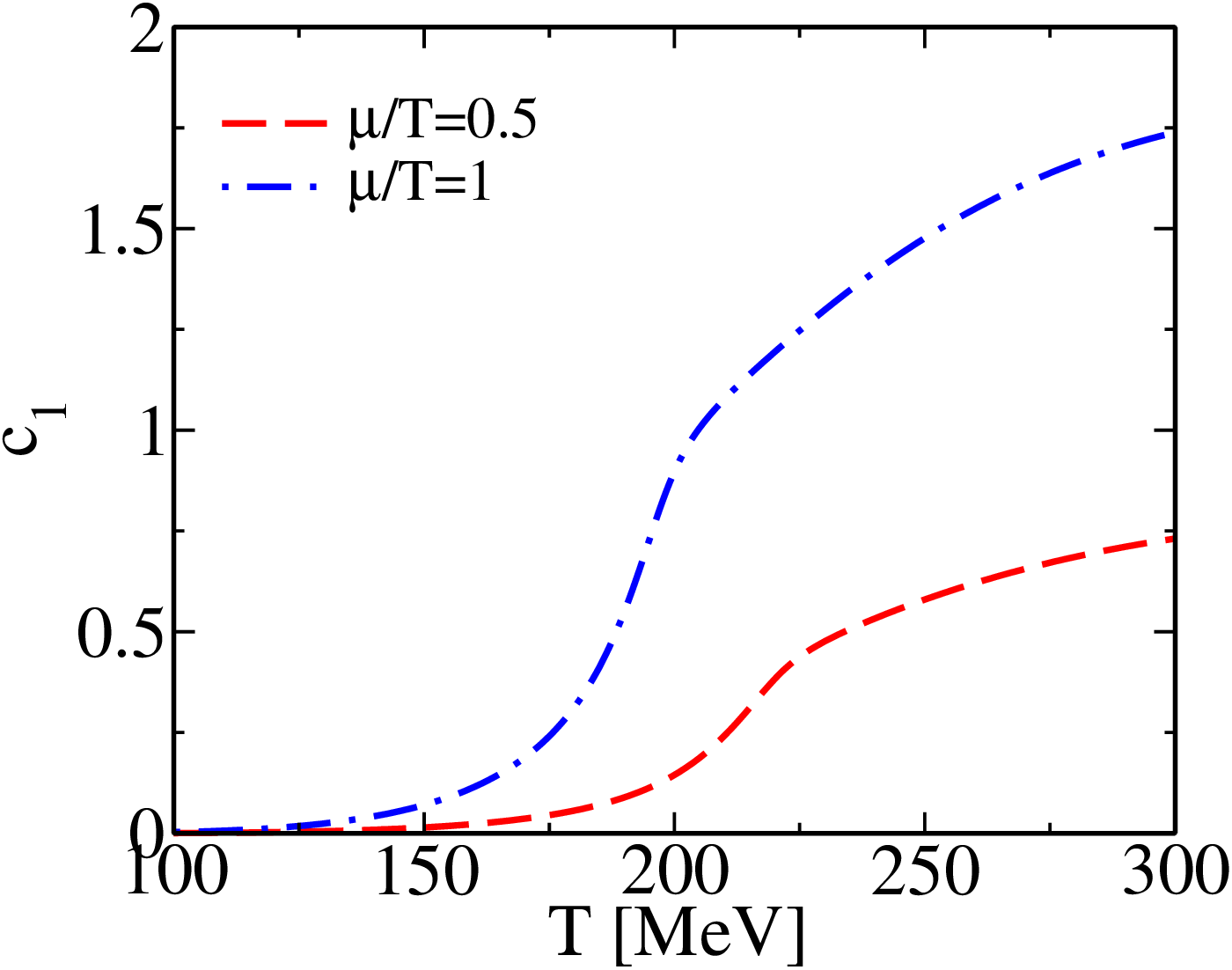}
  \includegraphics*[width=7.2cm]{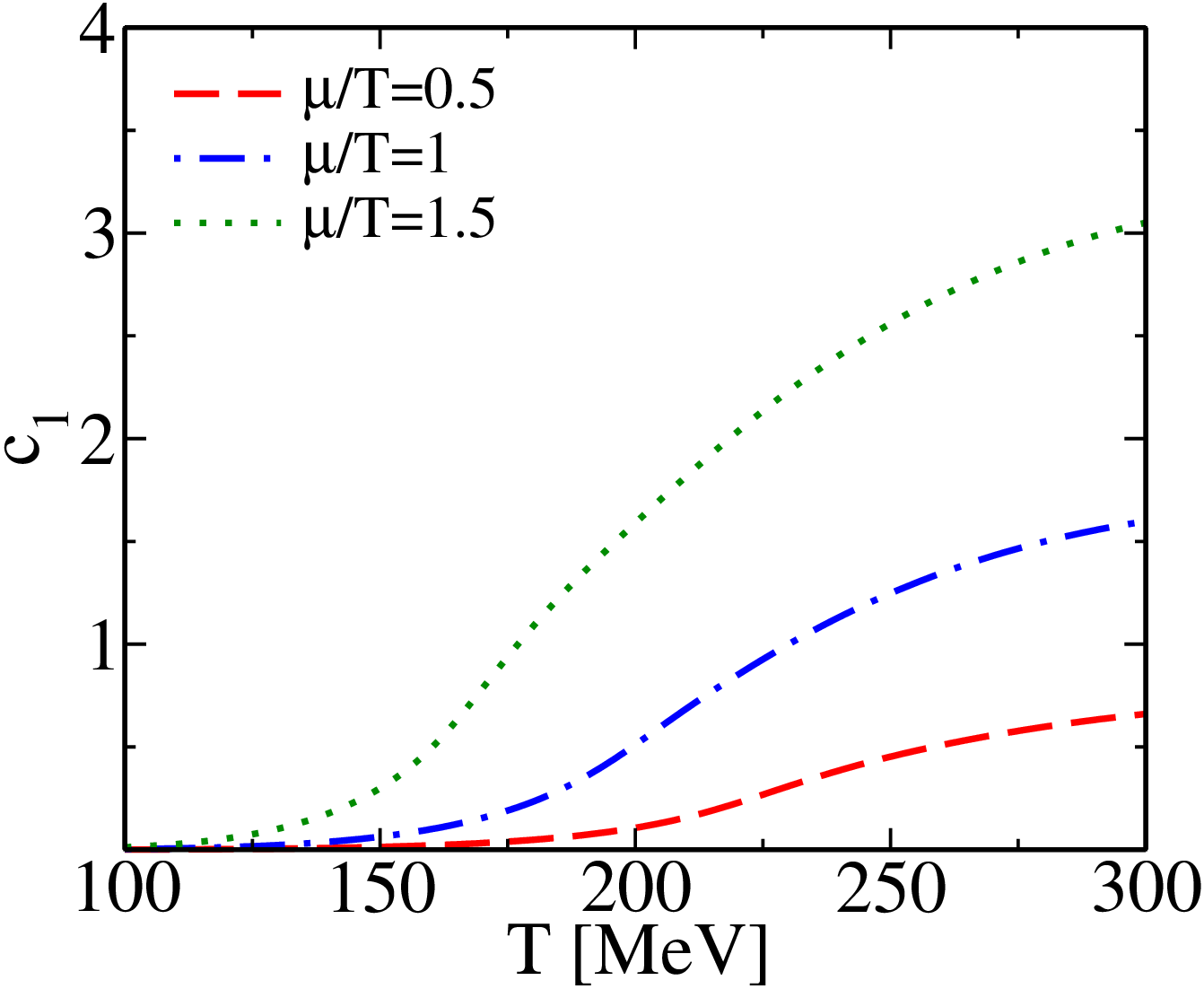}
  \caption {The baryon number density normalized by $T^3$, $c_1=n_q/T^3$, as a
    function of temperature for different values of $\mu/T$ for the {
      PQM} model in the mean-field approximation (left panel) and in
    the FRG approach (right panel).}
  \label{fig:c1}
\end{figure*}


The minimum of the thermodynamic potential is determined by the
stationarity condition
\begin{equation}
  \left. \frac{d \Omega_k}{ d \sigma} \right|_{\sigma=\sigma_k}=\left. \frac{d
      \overline{\Omega}_k}{ d \sigma} \right|_{\sigma=\sigma_k} - c =0.
  \label{eom_sigma}
\end{equation}

The flow equation~(\ref{eq:frg_flow}) is solved numerically with the
initial cutoff $\Lambda=1.2$ GeV (see details in
Ref.~\cite{Skokov:2010wb}).  The initial conditions for the flow are
chosen to reproduce { the  vacuum properties:} the physical pion mass $m_{\pi}=138$
MeV, the pion decay constant $f_{\pi}=93$ MeV, the sigma mass
$m_{\sigma}=600$ MeV, and the constituent quark mass $m_q=300$ MeV at
the scale $k=0$.  The symmetry breaking term, $c=m_\pi^2 f_\pi$,
corresponds to an external field and consequently does not flow. In
this work we neglect the flow of the Yukawa coupling $g$, {which is
not expected to be significant for the present studies~(see e.g. Refs.~\cite{Jungnickel,Palhares:2008yq}). }

By solving the equation~(\ref{eq:frg_flow}) one gets thermodynamic
potential for quarks and mesonic subsystems, $\Omega_{k\to0} (\ell,
\ell^*;T, \mu)$, as the function of the Polyakov loop variables $\ell$
and $\ell^*$. The full thermodynamic potential $\Omega(\ell, \ell^*;T,
\mu)$ in the PQM model which includes quarks, mesons and gluons
degrees of freedom is obtained by adding to $\Omega_{k\to0} (\ell,
\ell^*;T, \mu)$ the effective gluon potential ${\cal U}(\ell,
\ell^*)$:
\begin{equation}
  \Omega(\ell, \ell^*;T, \mu) = \Omega_{k\to0} (\ell, \ell^*;T, \mu) + {\cal U}(\ell, \ell^*).
  \label{omega_final}
\end{equation}
At a given temperature and chemical potential, the Polyakov loop
variables, $\ell$ and $\ell^*$, are determined by the stationarity
conditions:
\begin{eqnarray}
  \label{eom_for_PL_l}
  &&\frac{ \partial   }{\partial \ell} \Omega(\ell, \ell^*;T, \mu)  =0, \\
  &&\frac{ \partial   }{\partial \ell^*}  \Omega(\ell, \ell^*;T, \mu)   =0.
  \label{eom_for_PL_ls}
\end{eqnarray}

The thermodynamic potential~(\ref{omega_final}) does not contain
contributions of statistical modes with momenta larger than the cutoff
$\Lambda$.  In order to obtain the correct high-temperature behavior
of thermodynamic functions we need to supplement the FRG potential with the
contribution of high-momentum states.  A simplified model for
implementing such states was proposed in Ref.~\cite{Braun:2003ii}
where the contributions of the $k > \Lambda$ states to the flow is
approximated by that of a non-interacting gas of quarks and gluons.
For the PQM model we generalize this procedure by including for $k >
\Lambda$ the flow equation of interacting quarks with Polyakov loops~\cite{Skokov:2010wb}
\begin{eqnarray}\label{eq:qcdflow}
  \del_k \Omega_k^{\Lambda}(T,\mu)&=&-\frac{N_c N_f k^3}{3\pi^2}
  \\
  && \hspace*{-1cm} \Big[ 1-
  N(\ell,\ell^*;T,\mu)-\bar{N}(\ell,\ell^*;T,\mu)\Big],\nonumber
\end{eqnarray}
where the dynamical quark mass is neglected.  In addition, since the
effective gluon potential ${\cal U}(\ell, \ell^*)$ is fitted to
reproduce the Stefan-Boltzmann limit at high temperatures, the explicit
gluon contribution is omitted for consistency.

To obtain the complete thermodynamic potential of the PQM model we
integrate Eq.~(\ref{eq:qcdflow}) from $k=\infty$ to $k=\Lambda$ where
we switch to the PQM flow equation (\ref{eq:frg_flow}).  Divergent
terms in the high-momentum flow equation (\ref{eq:qcdflow}) are
independent of mesonic and gluonic fields as well as of temperature
and chemical potential. Consequently, such terms can be absorbed to an
unobservable constant shift of the thermodynamic potential.


%

%

\begin{figure*}
  \includegraphics*[width=7.5cm]{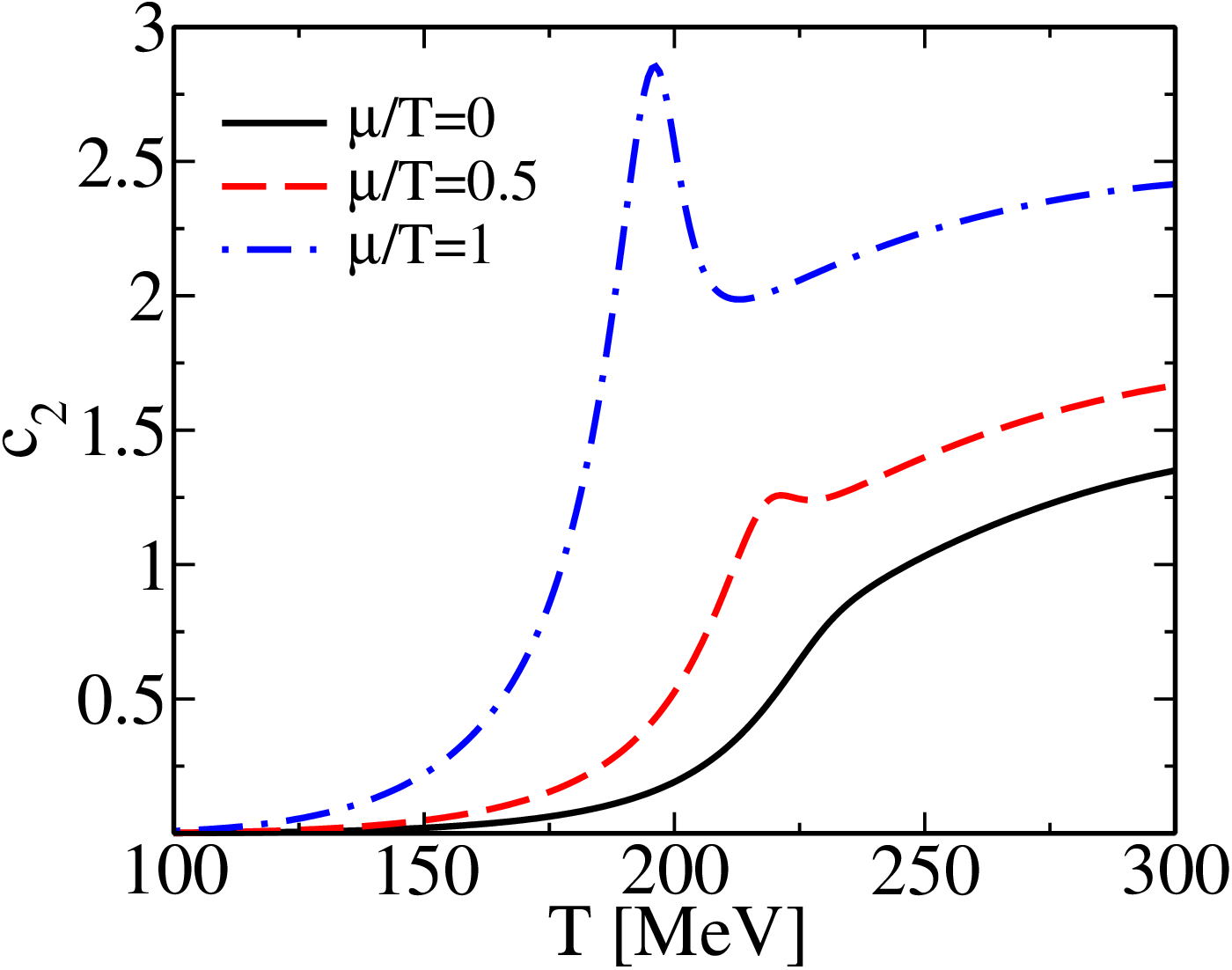}
  \includegraphics*[width=7.2cm]{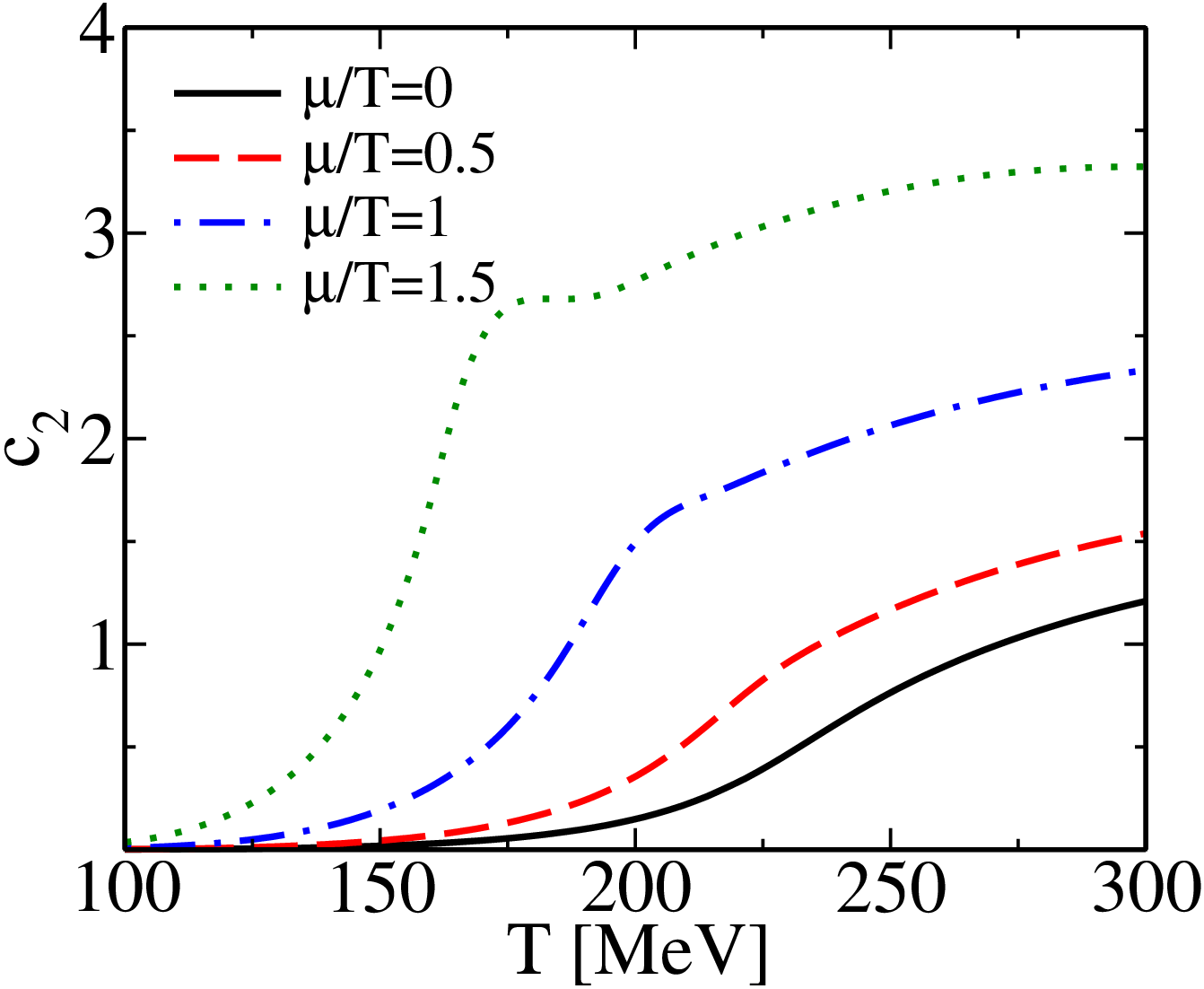}
  \caption{ The coefficient $c_2$ as a function of temperature for
    different values of $\mu/T$ for the { PQM} model in the mean-field
    approximation (left panel) and in the FRG approach (right panel).
  }
  \label{fig:c2}
\end{figure*}

\subsection{The mean-field approximation} \label{sec:mf}

To illustrate the importance of mesonic fluctuations
on thermodynamics in the PQM model we will compare
the FRG results with those obtained under the mean-field
approximation. In the latter, both quantum and thermal fluctuations
are neglected and the mesonic fields are replaced by their classical
expectation values.

\begin{widetext}

  In the PQM model the thermodynamical potential derived under the
  mean-field approximation reads~\cite{Schaefer:PQM}:
  \begin{equation}
    \Omega_{MF} = {\cal U}(\ell,\ell^*) + U(\langle\sigma\rangle, \langle\pi\rangle=0) + \Omega_{q\bar{q}} (\langle\sigma\rangle,\ell,\ell^*).
    \label{Omega_MF}
  \end{equation}
  Here, the contribution of quarks with dynamical mass
  $m_q=g\langle\sigma\rangle$ is given by
  \begin{equation}
    \Omega_{q\bar{q}} (\langle\sigma\rangle, \ell,\ell^*) = - 2 N_f T \int \frac{d^3 p}{(2\pi)^3} \left\{
      \frac{N_c E_q}{T} 
      + \ln
      g^{(+)}(\langle\sigma\rangle, \ell, \ell^*; T, \mu) +  \ln
      g^{(-)}(\langle\sigma\rangle,\ell, \ell^*; T, \mu) \right\},
    \label{Omega_MF_q}
  \end{equation}
  where
  \begin{eqnarray}
    \label{g}
    g^{(+)}(\langle\sigma\rangle,\ell, \ell^*; T, \mu) &=& 1 + 3 \ell
    \exp[-(E_q-\mu)/T] + 3 \ell^*\exp[-2(E_q-\mu)/T] + \exp[-3(E_q-\mu)/T], \\
    g^{(-)}(\langle\sigma\rangle,\ell, \ell^*; T, \mu) &=& g^{(+)} (\langle\sigma\rangle,\ell^*, \ell; T, -\mu);
  \end{eqnarray}
  and $E_q = \sqrt{p^2+m_q^2}$ is the quark quasi-particle energy. The
  first term in Eq.~(\ref{Omega_MF_q}) is a divergent vacuum
  fluctuation contribution which has to be properly
  regularized. Following our previous studies~\cite{MFonVT} we use the
  dimensional regularization, which introduces an arbitrary scale
  $M$. The renormalization procedure allows to compensate the physics
  dependence on this scale. The finite contribution of
  the vacuum term to  Eq.~(\ref{Omega_MF_q}) reads~\cite{MFonVT}
  \begin{equation}
    \Omega_{q\bar{q}}^{vac}  =  - \frac{N_c N_f}{8 \pi^2} m_q^4 \ln\left(\frac{m_q}{M}\right).
    \label{vacuum_term}
  \end{equation}
  The importance and influence of this contribution on the thermodynamics
  of chiral models was demonstrated and studied in detail in Refs. \cite{MFonVT} and
  \cite{Nakano:2009ps}.

\end{widetext}

The equations of motion for the mean fields are obtained by requiring
that the thermodynamic potential is stationary with respect to changes
of $\langle\sigma\rangle$, $\ell$ and $\ell^*$:
\begin{equation}
  \frac{\partial \Omega_{MF}}{\partial \langle\sigma\rangle} = \frac{\partial \Omega_{MF}}{\partial \ell} = \frac{\partial \Omega_{MF}}{\partial \ell^*} =0.
  \label{EOM_MF}
\end{equation}

The model parameters are fixed to reproduce the same vacuum physics as
in the FRG calculation.

\begin{figure*}
  \includegraphics*[width=7.5cm]{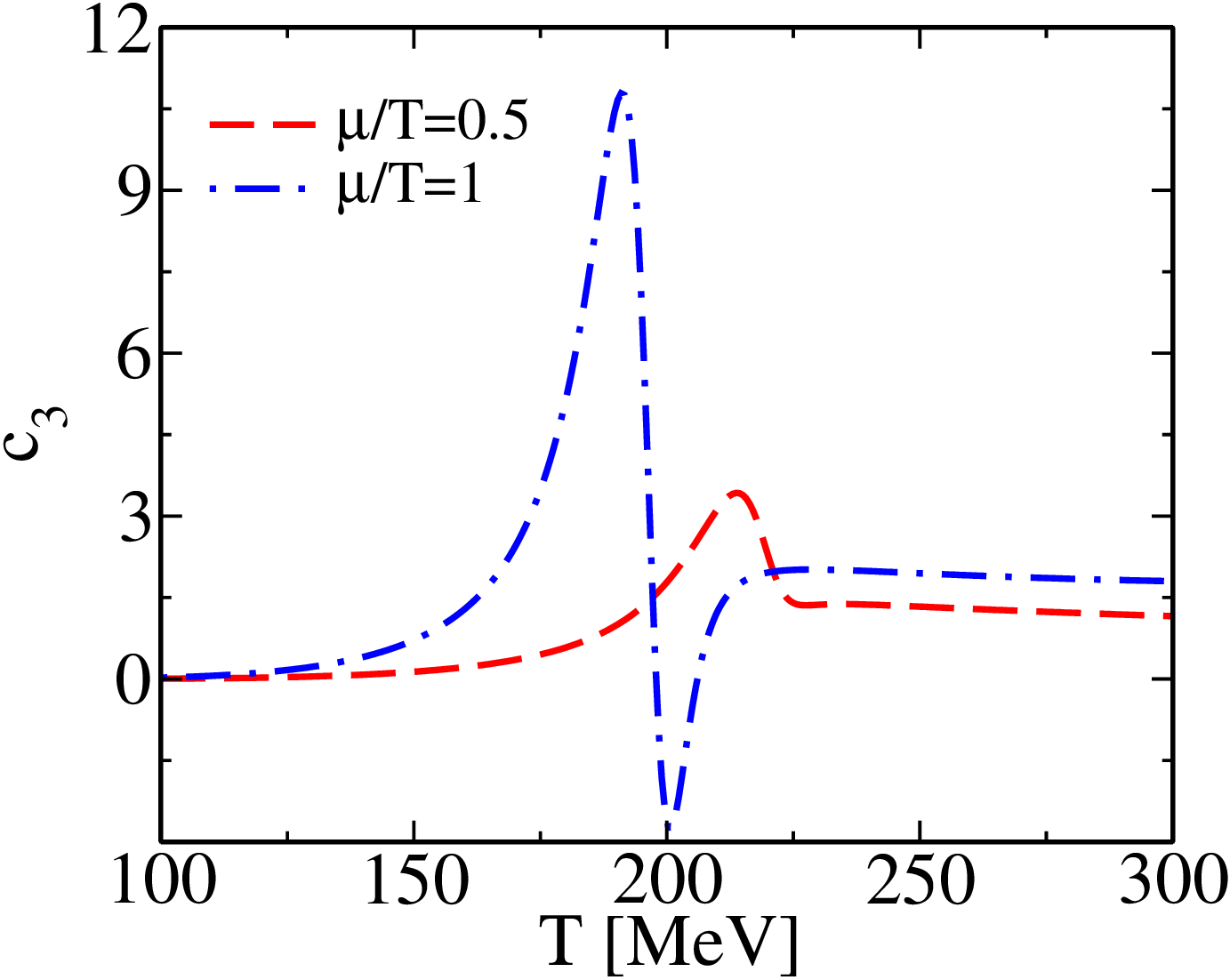}
  \includegraphics*[width=7.2cm]{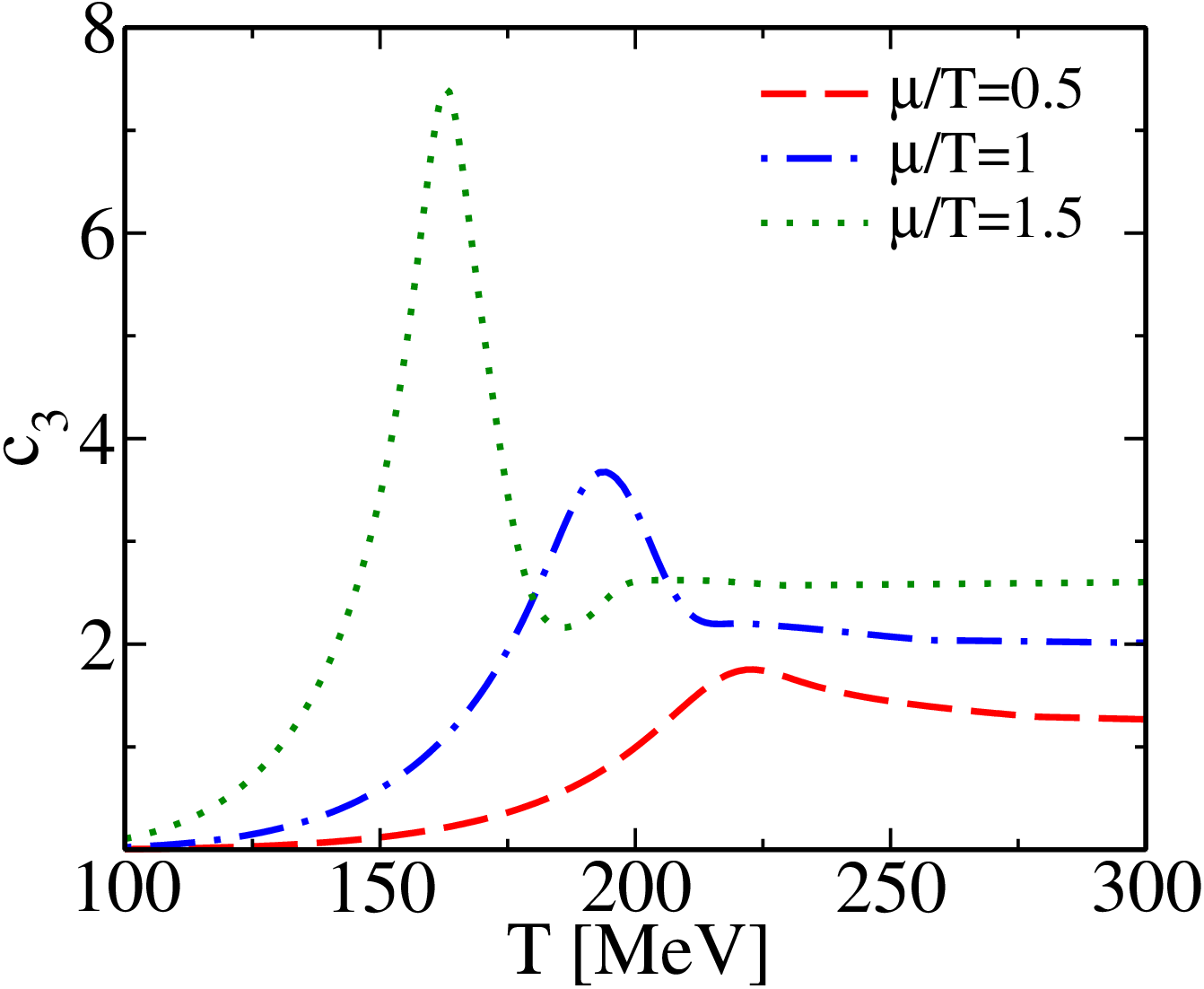}
  \caption {The coefficient $c_3$ as a function of temperature for
    different values of $\mu/T$ for the { PQM} model in the mean-field
    approximation (left panel) and in the FRG approach (right
    panel). }
  \label{fig:c3}
\end{figure*}

\section{Thermo\-dynamics of the PQM model}\label{sec:thermo}

The thermodynamic potential introduced in Eqs. (\ref{omega_final}) and
(\ref{Omega_MF}) provides basis to explore critical properties of the
PQM model at finite baryon density within the FRG approach and under the mean-field  approximation.

To find the potential at finite temperature and chemical potential
one needs to solve the FRG flow equation (\ref{eq:frg_flow}).
{ This equation is solved by  numerical methods based on }the Taylor
series expansion~\cite{Litim:2002cf,Skokov:2010wb}. This method is successful in studying
thermodynamics at finite density and temperature~\cite{SFR,Nakano:2009ps,Skokov:2010wb}
in the regime where the system exhibits the crossover or the second-order
 phase transition.  For the solution of the FRG flow equations in
the regime of the first-order phase transition, where the
thermodynamical potential develops two minimums, different
numerical methods are required~\cite{Adams:1995cv,Nakano:2009ps}. In
the present work we restrict our considerations only to the parameter range
where the PQM model exhibits the crossover or the second-order chiral
phase transitions.

\subsection{The phase diagram}

The PQM model is expected to  belong to the same universality class as QCD
and thus should
exhibit a generic phase diagram with the critical point at
non-vanishing chemical potential. In the chiral limit the phase
diagram is identified by divergences of the chiral susceptibility. At
finite quark mass the chiral transition is of crossover type. In this case
the pseudocritical temperature { and chemical potential are located by determining  the
maximum} of the chiral susceptibility or the temperature derivative of
the chiral order parameter. The position of the CEP is identified by the
properties of the sigma mass.
The
temperature and chemical potential where the sigma mass vanishes
correspond to the position of the CEP. One can equivalently consider
the net-quark number fluctuations to identify the critical point which
according to $Z(2)$ universality diverge at the CEP.

Fig.~\ref{fig:PD} shows the phase diagrams of the PQM model obtained
in the FRG approach and in the mean-field approximation. For the
physical pion mass and moderate values of the chemical potential the
PQM model exhibits a smooth crossover chiral transition.  In
Fig.~\ref{fig:PD}
we define the transition region as bands where the temperature
derivative  of the order parameter exhibits $5\%$-deviations from its
maximal value. At larger chemical potentials the crossover line terminates at the
CEP where the transition is of the second order
and belongs to the universality class of { the three dimensional
Ising model.} 

Comparing the resulting phase diagrams of the PQM model obtained
within the mean-field approximation  and FRG approach we find a clear shift of the position of
the chiral phase boundary to higher temperatures due to mesonic
fluctuations.  Such shifts were previously found in the QM model
within the FRG approach~\cite{Schaefer:2006ds,Nakano:2009ps}. However, in our
studies due to the gluonic background, which is explicitly included in
our FRG calculations, we  also find a significant shift of the CEP to higher
temperature.

In the following we focus on observables that are related to the
non-vanishing
net-quark number density and consider moments of  quark number fluctuations.
We discuss how such fluctuations depend on the quark chemical
potential in the presence of the gluonic background within the FRG approach.


\begin{figure*}
  \includegraphics*[width=7.5cm]{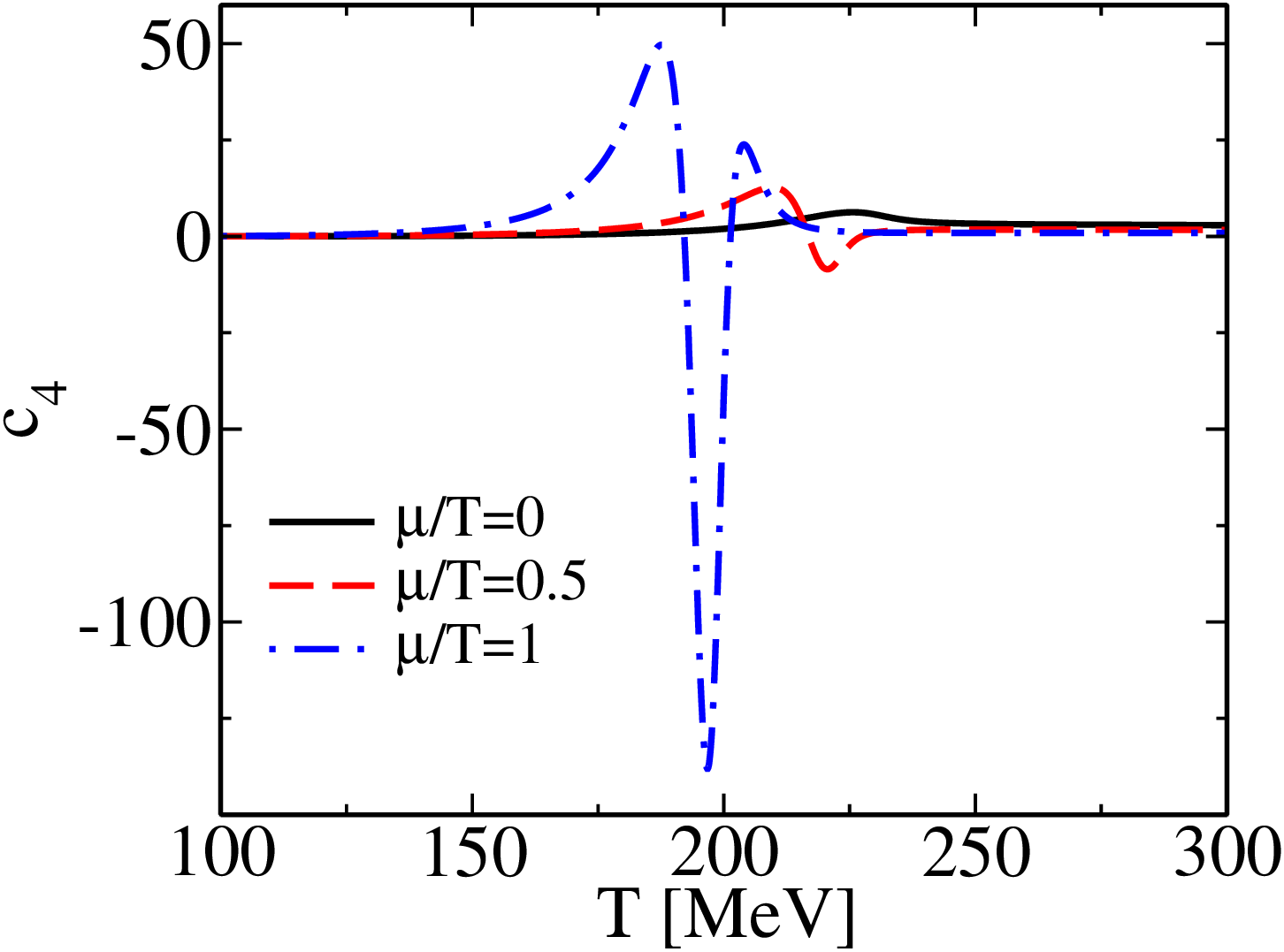}
  \includegraphics*[width=7.3cm]{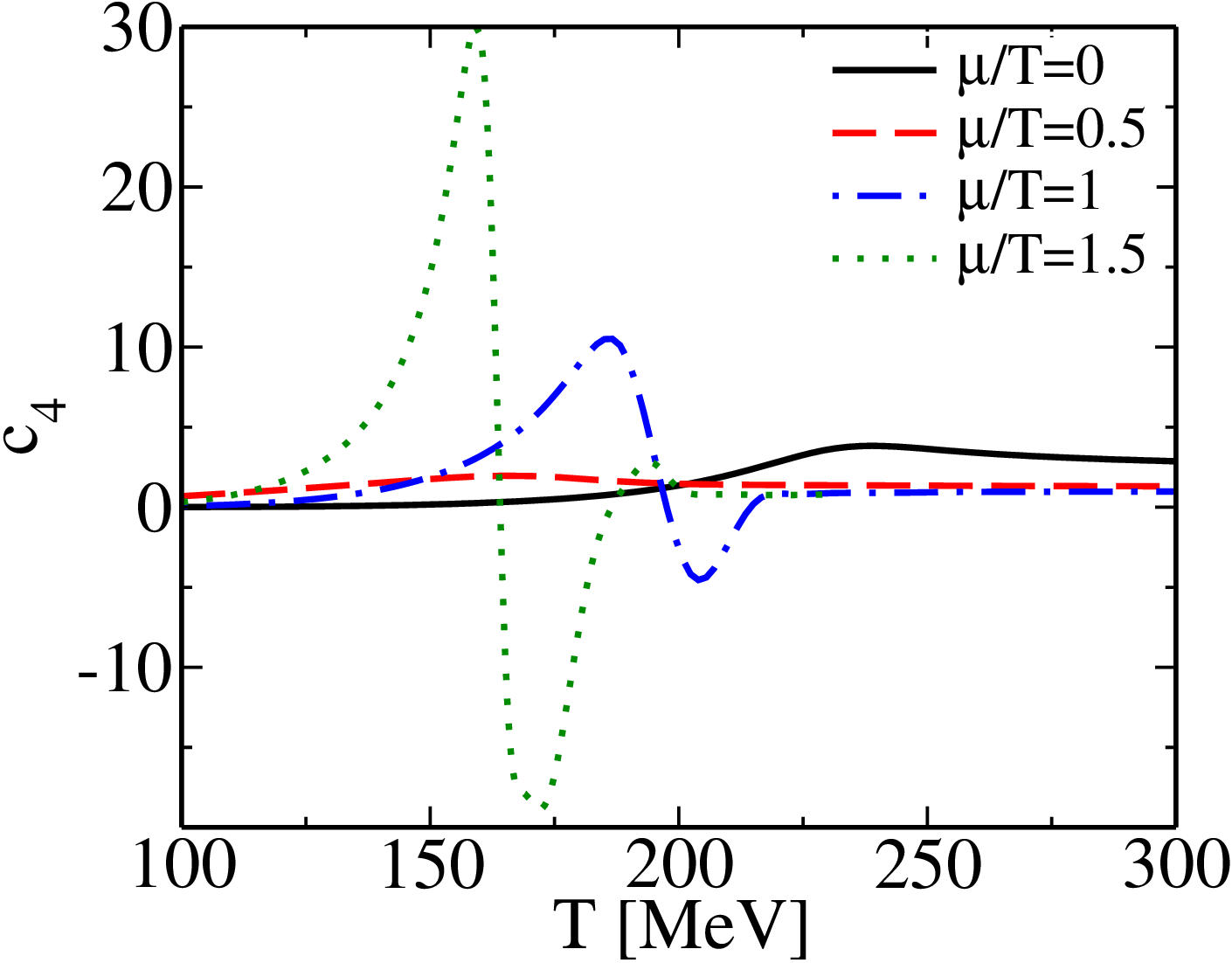}
  \caption{ The coefficient $c_4$ as a function of temperature for
    different values of $\mu/T$ for the { PQM} model in the mean-field
    approximation (left panel) and in the FRG approach (right panel).
  }
  \label{fig:c4}
\end{figure*}

%
%

\subsection{Quark number density fluctuations}

The fluctuations of conserved charges are observables that provide
direct information on critical properties related with the chiral
symmetry restoration. Fluctuations related
with the baryon number conservation are of particular interest.
In an equilibrium medium a divergence of the net-quark number
susceptibility is a direct evidence for the existence of the CEP~\cite{Stephanov:1999zu}. Consequently, any
non-monotonic dependence of these fluctuations on collision energy in
heavy-ion collisions was proposed as a phenomenological method to
verify the CEP~\cite{Stephanov:1999zu}. In a non-equilibrium system, the net-quark
number susceptibility was also shown to signal the first-order chiral
phase transition due to spinodal decomposition~\cite{CS}.

The fluctuations of the net-quark number density  are
characterized by the generalized susceptibilities,
\begin{equation}
  c_n(T)=\frac{\del^n[p\,(T,\mu)/T^4]}{\del(\mu/T)^n}.
\end{equation}

The first coefficient $c_1=n_q/T^3$
characterises  the response of the net-quark number density to changes
in the quark chemical potential. The second coefficient
\begin{equation}
  c_2 = {\frac{\chi_q} {T^2}}=  \frac{1}{V T^3}  \langle(\delta N_q)^2\rangle,
\end{equation}
with $\delta N_q= N_q-\langle N_q\rangle$ is proportional to the susceptibility  of the net-quark number density, $\chi_q$.
The third  and fourth order moments can be expressed through $\delta N_q$ as
\begin{eqnarray}
  c_3 &=& \frac{1}{V T^3} \langle(\delta N_q)^3\rangle, \\
  c_4 &=& \frac{1}{V T^3} (\langle(\delta N_q)^4\rangle-3\langle(\delta N_q)^2\rangle^2) \label{fluctuations}.
\end{eqnarray}

The coefficients $c_n(T)$ are sensitive probes of the chiral phase
transition. They indicate the position, the order and, in case of the
second-order phase transition, the universality class of corresponding phase transition.
The net-quark number density, $n_q$, is discontinuous at the first-order
transition, whereas the susceptibility $c_2$ and higher cumulants
diverge at the
CEP~\cite{Stephanov:2007fk,Stephanov:2008qz,Athanasiou:2010kw}.
In the chiral limit and at non-zero chemical potential, all generalized susceptibilities
$c_n(T)$ with $n>2$ diverge at the $O(4)$ chiral critical
line~\cite{Ejiri:2005wq}.  
Moreover, they also diverge  at the spinodal lines  of the first-order phase transition~\cite{CS}.

A very particular role is attributed to the so-called kurtosis of the
net-quark number fluctuations~\cite{Ejiri:2005wq,F1,kurtosis} which is
defined  as the ratio
\begin{equation}\label{eq:ratio_c42}
  R_{4,2}=\frac{c_4}{c_2}.
\end{equation}
This key observable is not only sensitive to the chiral but also to
the deconfinement transition. At vanishing chemical potential,  in the asymptotic regime of
high and low temperatures the
kurtosis reflects  quark content of the baryon-number carrying
effective degrees of freedom~\cite{Ejiri:2005wq,kurtosis}.  Therefore,
at low temperatures in the confined phase, $R_{4,2}\simeq N_q^2=9$ while
for high temperatures one recovers an ideal gas of quarks with $R_{4,2}\sim 1$~\footnote{More precisely, this
  number is $6/\pi^2$ due to quantum statistics.}.
\begin{figure*}
  \includegraphics*[width=7.5cm]{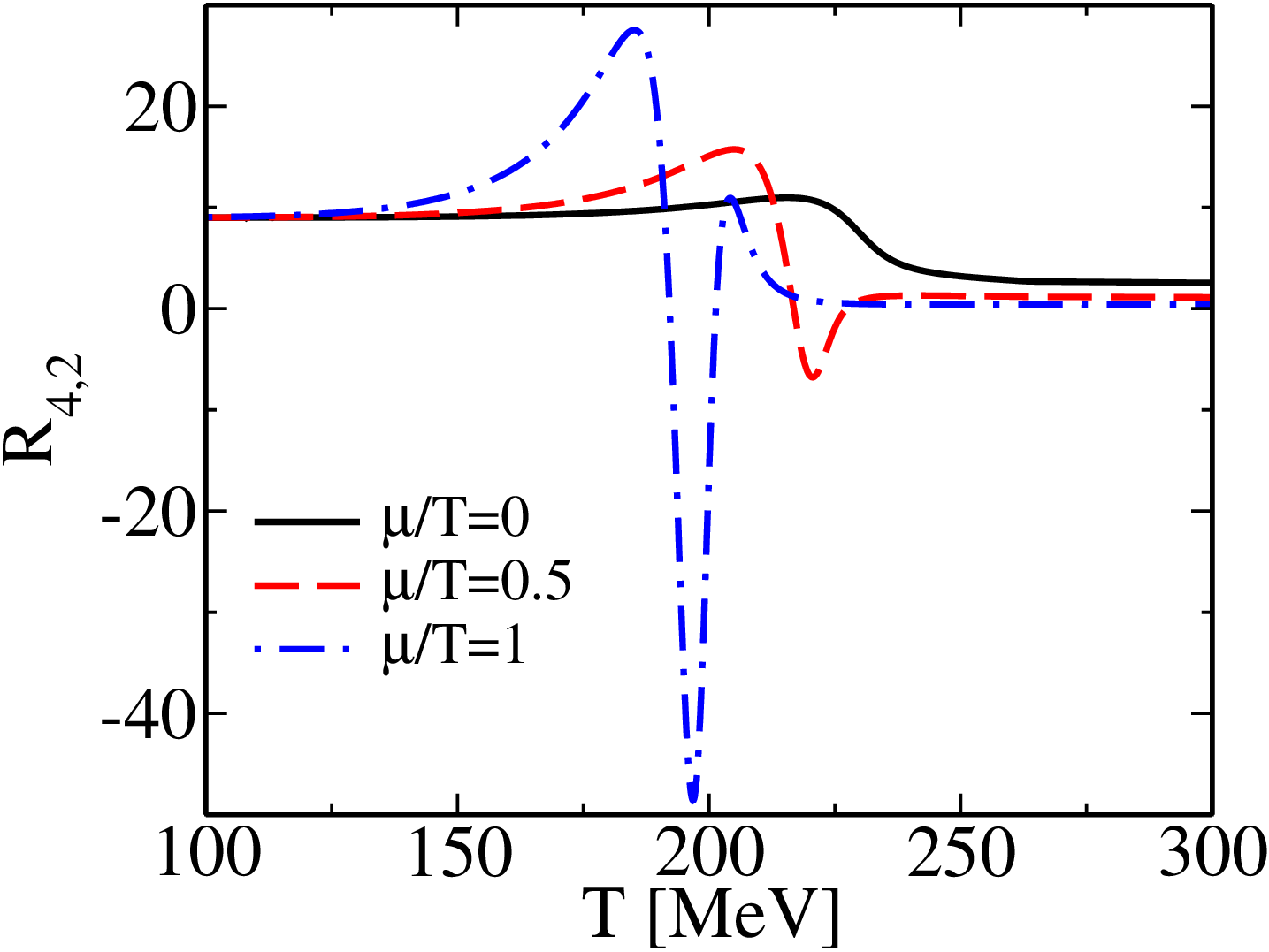}
  \includegraphics*[width=7.3cm]{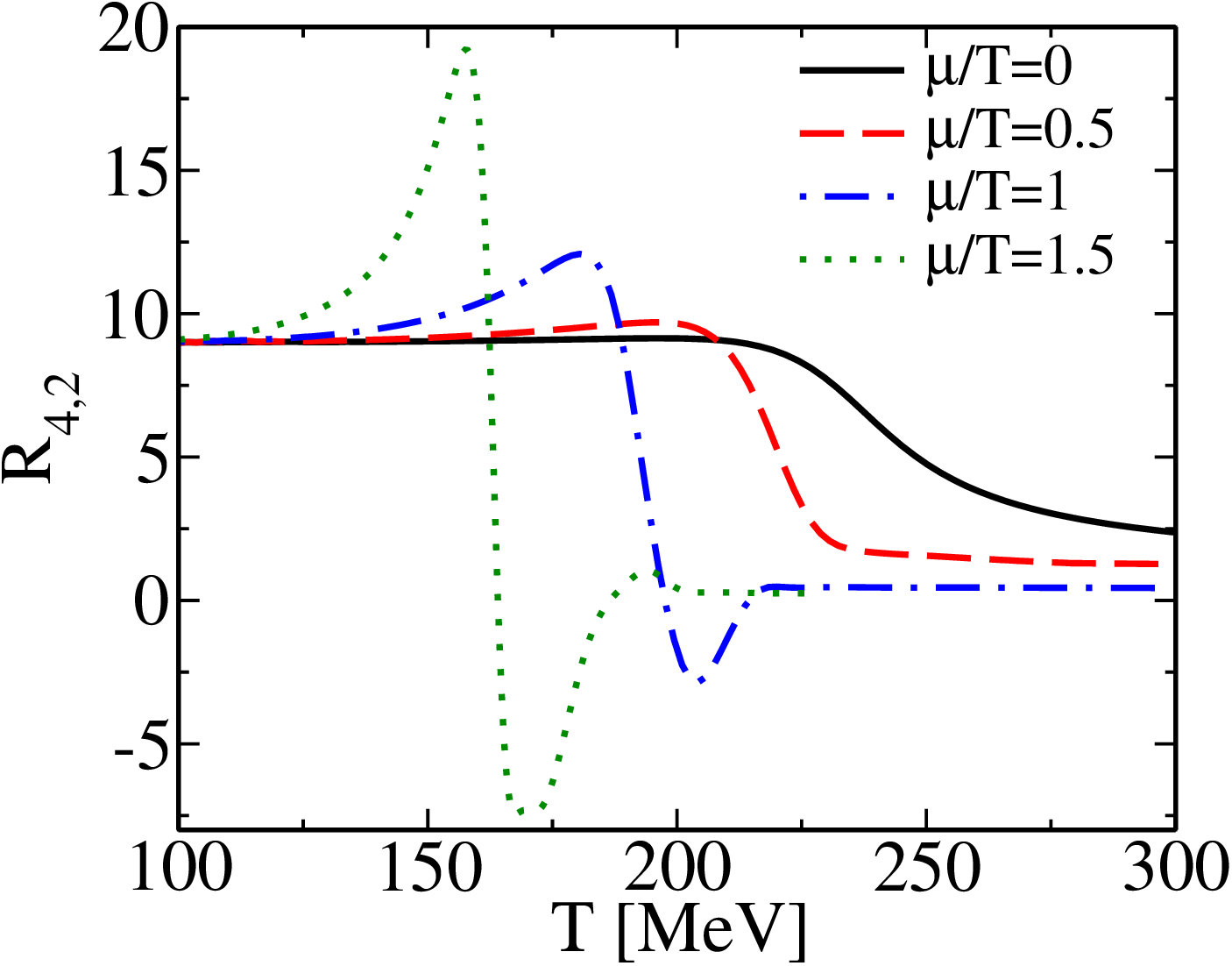}
  \caption{ The kurtosis $R_{4,2}$ as a function of temperature for
    different $\mu/T$ calculated in the { PQM} model under the
    mean-field approximation (left panel) and in the FRG approach
    (right panel).  }
  \label{fig:r}
\end{figure*}

The properties of different moments of the net-quark number
fluctuations in the presence of the chiral phase transition were
studied in the literature both in terms of the LQCD ~\cite{lgt2} as
well as in different models \cite{SFR,Fu:2009wy,Schaefer:2009ui,Skokov:2010wb,MFonVT,Karsch:2010ck}.
 In particular, the
importance of the quark susceptibility and the kurtosis as signatures
of the deconfinement and the chiral phase transition as well as the
CEP was discussed~\cite{kurtosis}. The influence and dependence of
these fluctuations on the quark mass was also analyzed on the lattice
and in effective
models~\cite{SFR,Fu:2009wy,Schaefer:2009ui,Skokov:2010wb,MFonVT}.
However, only little is known
about chemical potential dependence of the higher cumulants $c_n$, particularly
 with $n>2$.  Such dependence can be obtained in the PQM model
from the thermodynamic potential introduced in Eqs.~(\ref{omega_final}) and~(\ref{Omega_MF}).  In
Figs.~\ref{fig:c1}--\ref{fig:c4} we quantify the first four moments obtained in
the PQM model under the mean-field approximation  and in the FRG
approach for different values of the ratio $\mu/T$. The lines of the fixed ratio $\mu/T$
also indicated on the phase diagram in Fig.~\ref{fig:PD}.

Fig.~\ref{fig:c1} shows the net-quark number density normalized by
$T^3$, $c_1=n_q/T^3$, for different values of $\mu/T$.  The cumulant
$c_1$ as well as all $c_{2n+1}$, for $n=1, 2, 3, \ldots$,
are vanishing at zero chemical potential, $\mu=0$. At finite $\mu$
and in the chiraly broken phase the net-quark number, $n_q$,  is strongly increasing
function of $\mu/T$. In the low temperature phase due to the large dynamical quark mass the
$n_q$ increases approximately as $\sinh(3\mu/T)$. This is a direct
consequence of the "statistical confinement" properties of the PQM
model which for small expectation values of Polyakov loops  $l\ll1$
suppresses the single and double quark modes in the partition function
 as seen from Eqs.~(\ref{n1}) and~(\ref{g}).  At
high temperatures/chemical potentials $n_q$ behaves as polynomial in $\mu/T$. For a fixed
$\mu/T$ and in the pseudo-critical region where the chiral symmetry is
partially restored there is a clear change in the temperature dependence of
$n_q$.  At the pseudo-critical temperature the density exhibits a
kink-like structure which is particularly evident in the mean-field
calculations at larger $\mu/T$. In the FRG calculations and at the
corresponding $\mu/T$ the above kink-like structure in the density $n_q$ is
suppressed.  Thus, the meson fluctuations quantified by the FRG method
provide smoothing of the net-quark density near the crossover
transition.

\begin{figure*}
  \includegraphics*[width=7.5cm]{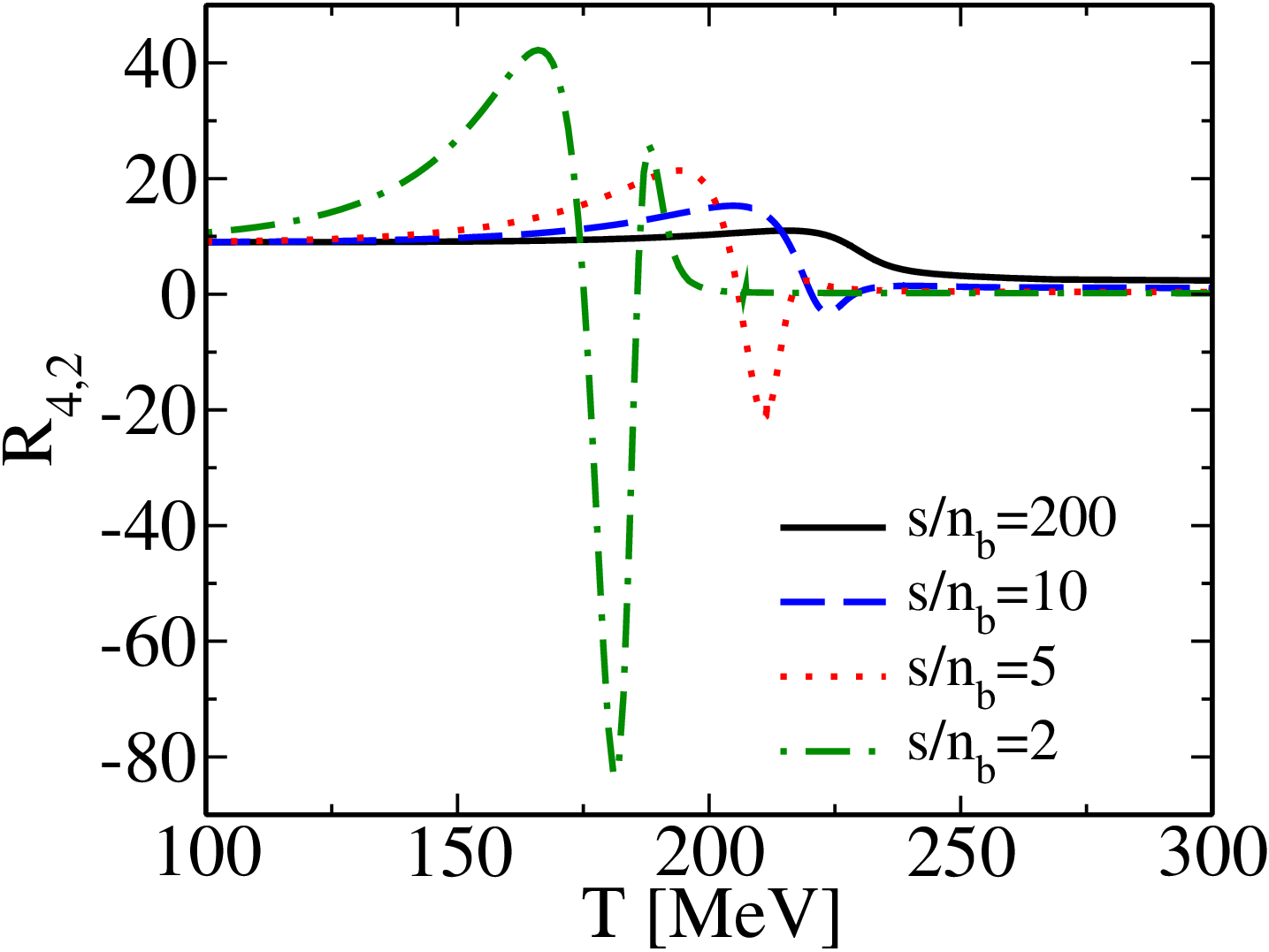}
  \includegraphics*[width=7.3cm]{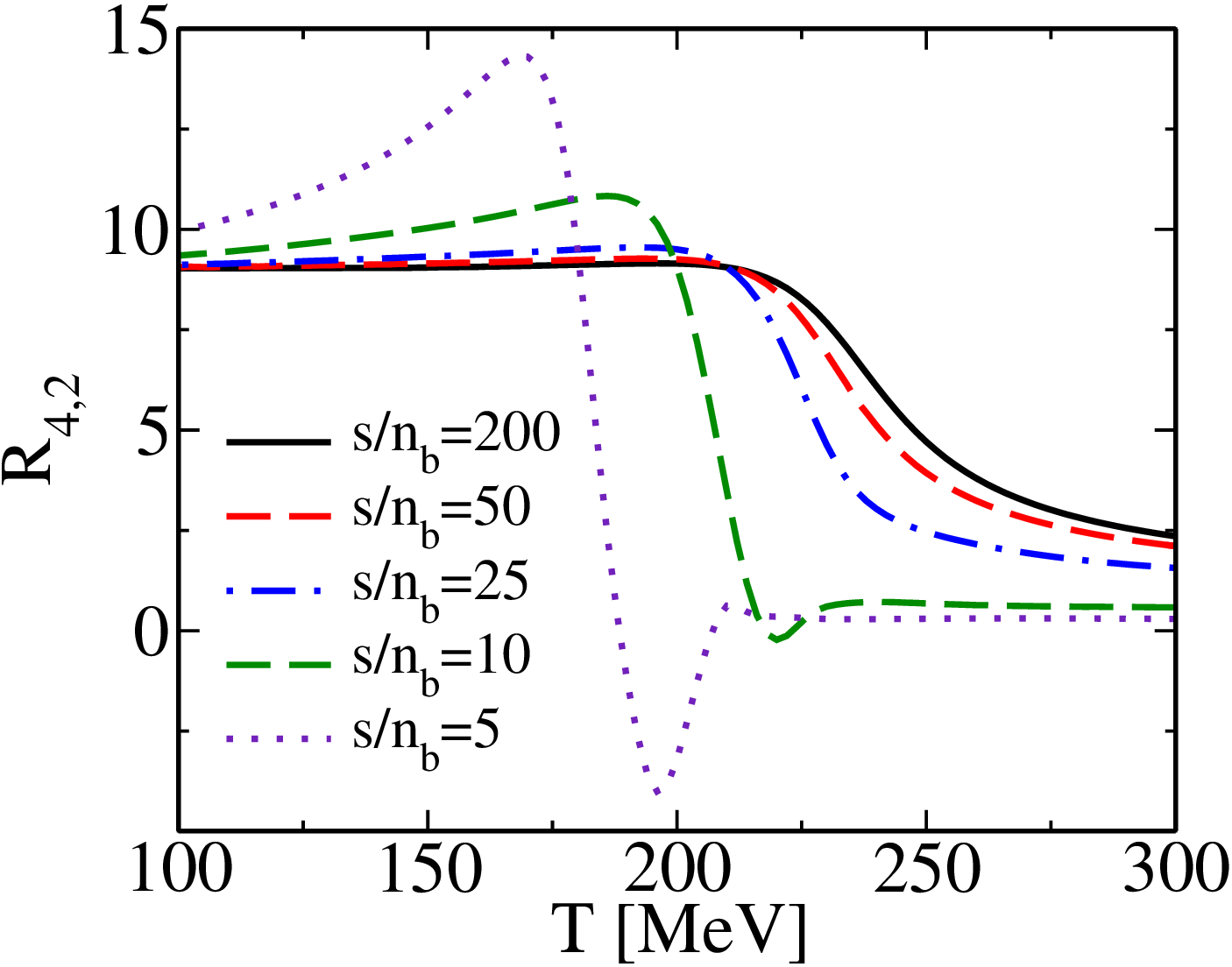}
  \caption{ The temperature dependence of the kurtosis $R_{4,2}$
    calculated in the PQM model at fixed values of the entropy density to net-quark density
    ($s/n_q$) under the mean-field approximation (left panel) and in
    the FRG approach (right panel).  }
  \label{fig:ris}
\end{figure*}

The appearance of the crossover chiral transition in the PQM model is
also transparent when considering the structure of isentropic
trajectories.  Because of  the entropy and baryon-number conservation such
trajectories  correspond to contours of the constant entropy density
per net-quark number density, $s/n_q$, in the
temperature-chemical potential plane. Fig.~\ref{fig:PD} shows
isentropes for different $s/n_q$ in the PQM model obtained under the
mean-field approximation  and in the FRG calculations. There is a clear
change  of slopes of isentropes along the transition line indicating
the change of the equation of state. The qualitative behavior of contours of constant $s/n_q$ obtained
in the PQM model is similar to that calculated previously within the
QM model~\cite{Nakano:2009ps}. In particular, the isentropes remain smooth when
the effect of long-wavelength meson fluctuations is consistently
included in the presence of the gluon background. There are also no
indications of any focusing towards the CEP in the PQM model.


The influence of the finite chemical potential on  $c_2$ (which is
proportional to the net-quark number susceptibility $c_2=\chi_q/T^2$)
is shown in Fig.~\ref{fig:c2} for the mean-field approximation and the
FRG approach.  At vanishing chemical potential the cumulant  $c_2$
increases monotonously with temperature.
  However, at finite chemical potential, the susceptibility
$c_2$ develops a peak structure.  The amplitude of this peak increases
with the chemical potential towards the CEP where $c_2$ diverges. In the
high temperature/chemical potential phase  $c_2$ converges to the Stefan Boltzmann limit
\begin{eqnarray}
  c_{2}^{ SB} &=& \frac{N_cN_f}{3} \left[ 1 + \frac{3 }{\pi^2} \left( \frac{\mu}{T}\right)^2 \right].
  \label{SB}
\end{eqnarray}

As it is seen in Fig.~\ref{fig:c2} the peak structure in $c_2$ is more
pronounced in the mean-field approximation at $\mu/T=1$ than in the FRG at
$\mu/T=1.5$. This is in spite of the fact that the location of the CEP
in the FRG is closer to the line of $\mu/T=1.5$ than the corresponding one for
the mean-field approximation to $\mu/T=1$ (see Fig. \ref{fig:PD}).
This shows that the criticality of  $c_2$ as a function of the
distance to the CEP appears earlier in the mean-field approximation
than in the FRG approach. This result is in agreement with the previous studies
in the QM model showing that the critical region shrinks due to
mesonic fluctuations~\cite{Schaefer:2006ds}.

The cumulants  $c_1$ and $c_2$ are sensitive to
changes in the chemical potential and are influenced by the meson
fluctuations and the gluon background. However, both these
coefficients are
 positive for all values of $\mu$ and $T$.  At
finite chemical potential the positivity of the first two coefficients is not
preserved for $c_n$-moments with $n>2$.  Figs.~\ref{fig:c3}
and~\ref{fig:c4} show the third and the fourth order cumulants 
of
the net-quark number density for different values of $\mu/T$. For a
large $\mu/T$ both these susceptibilities exhibit  a peculiar 
structure in the transition region. There is a minimum of $c_3$
which can reach negative value. The amplitudes of this minimum is
strongly suppressed by meson fluctuations in FRG approach.


The fourth order cumulant is strictly positive for vanishing chemical
potential. However, for higher values of $\mu/T$,  $c_4$ becomes
negative in the vicinity of the crossover transition. This
manifests the broadening of the net-quark number distribution in
comparison to the Gaussian one. Large values of $c_4$ in the broken
phase infer that the distribution becomes narrower than the
Gaussian. The chemical potential independent Stefan-Boltzmann limit $
c_{4}^{ SB} ={2 N_cN_f}/{\pi^2}$ is reproduced at temperatures
$T\gg T_c$.

\begin{figure*}
  \includegraphics*[width=7.5cm]{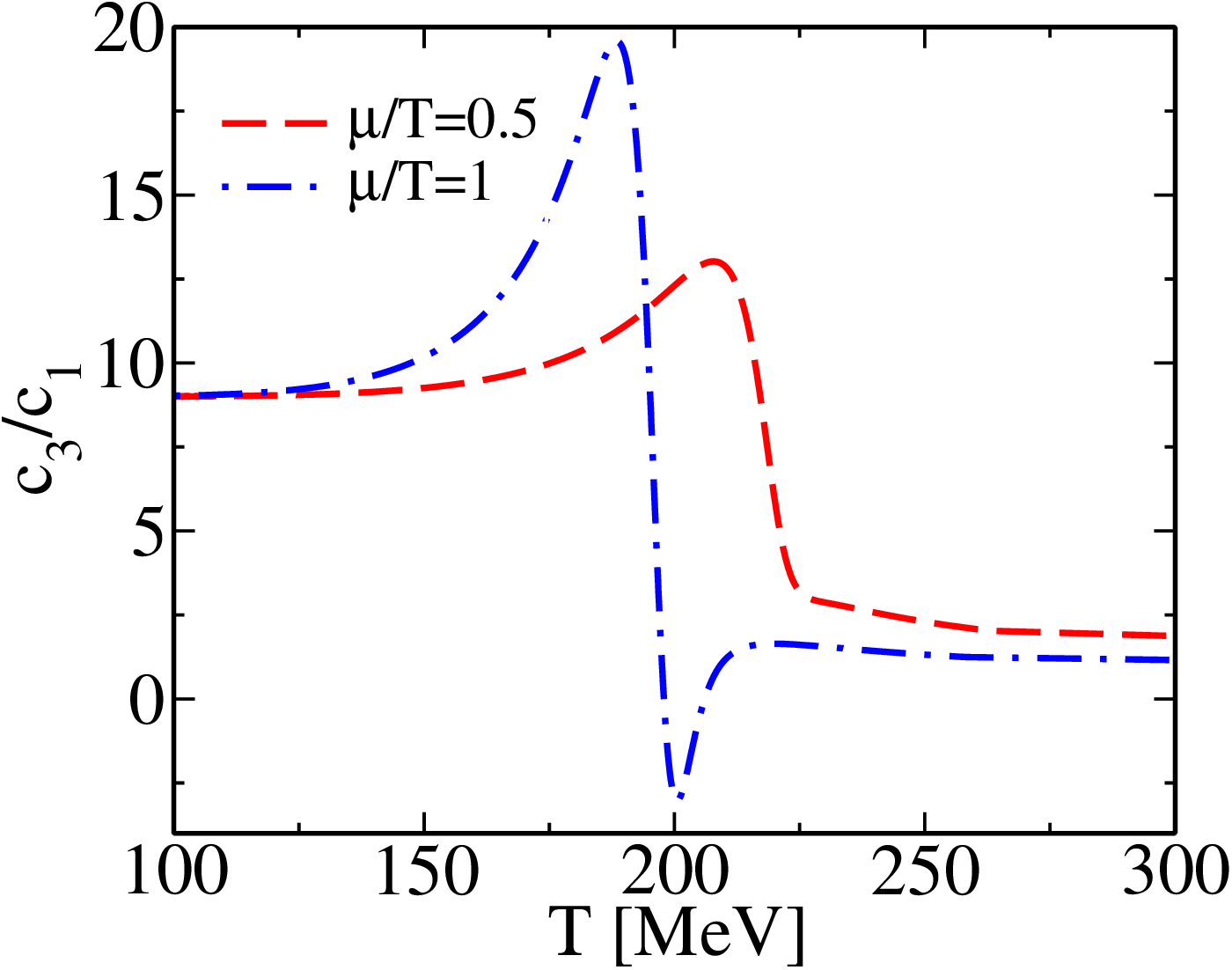}
  \includegraphics*[width=7.2cm]{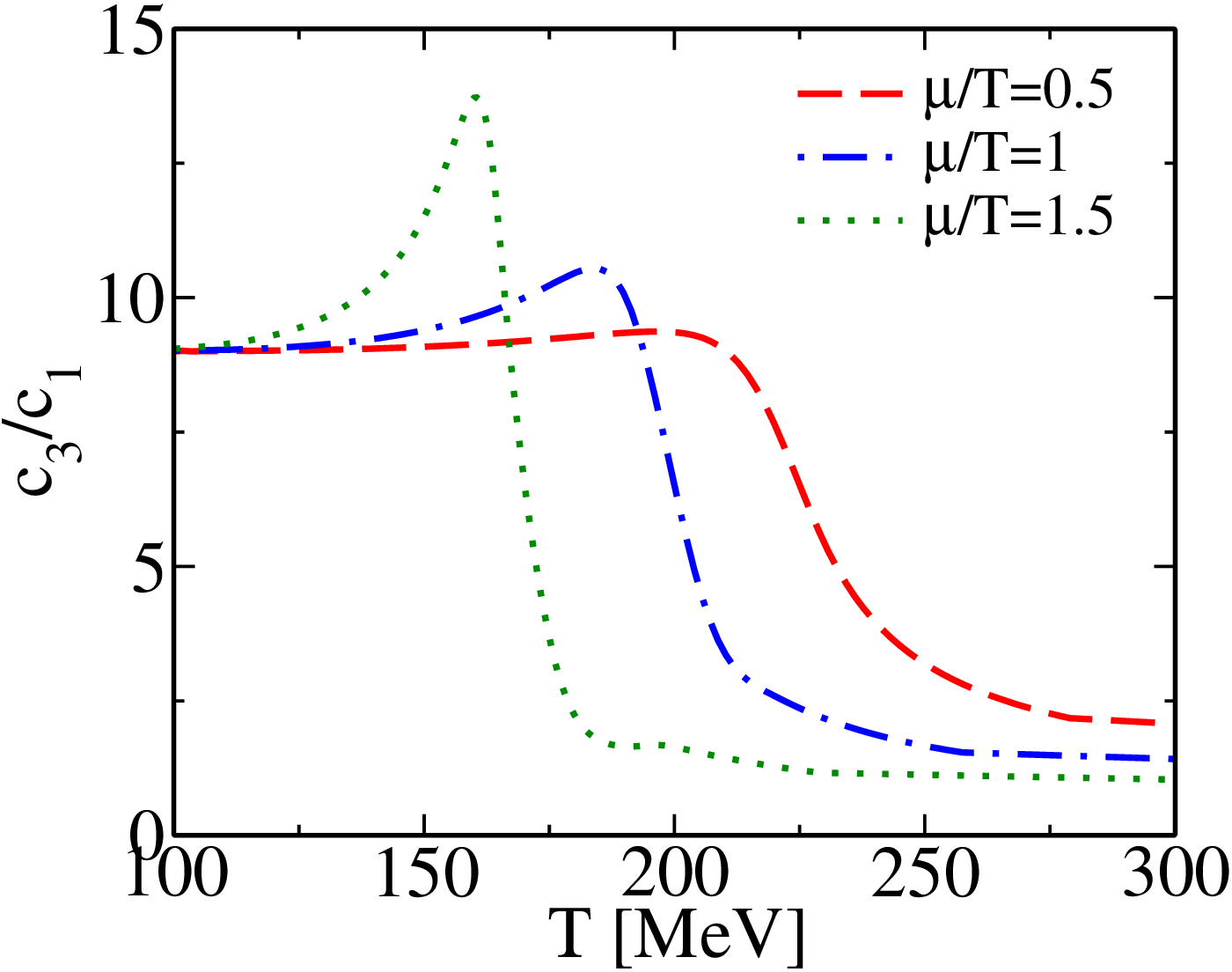}
  \caption {The ratio of $c_3$ to $c_1$ as a function of temperature
    for different values of $\mu/T$ obtained in the { PQM} model under
    the mean-field approximation (left panel) and in the FRG approach
    (right panel). }
  \label{fig:c3c1}
\end{figure*}

Comparing the mean-field with the FRG results for both $c_3$ and $c_4$
we conclude that mesonic  fluctuations  essentially modify properties
of different generalized quark susceptibilities. In the transition region $c_3$ and
$c_4$ are suppressed in the FRG relative to the mean-field  results.
{ Thus, the mean-field  approach  is only an approximate method to describe static thermodynamics
near the chiral phase transition.}

%
%

We have already indicated that the ratios of different $c_n$ are
sensitive to the deconfinement and the chiral phase transitions.
Figs.~\ref{fig:r} and~\ref{fig:ris} show the kurtosis
$R_{4,2}=c_4/c_2$ calculated as a function of temperature along
different paths in the temperature--chemical potential plane quantified by fixed $\mu/T$
and  $s/n_q$.  In the PQM model the kurtosis drops from $R_{4,2}\simeq 9$ to
$R_{4,2} \sim 1$ in the transition region. Such a change is explicitly
attributed to the change in quark content of baryon number carrying effective
degrees of freedom~\cite{Ejiri:2005wq}. In the PQM model and in the
chiraly broken phase the effective three quark states dominate, while
at high temperatures single quarks prevail. In the mean-field
approximation the kurtosis exhibits a peak at the transition
temperature at $\mu=0$. The height of these peaks depends not only on
the pion mass~\cite{kurtosis,karsch,Skokov:2010wb,MFonVT} but also
on the value of the chemical potential. The mesonic
fluctuations weaken the peak structure both at finite and at vanishing
quark density. For finite chemical potential  the kurtosis becomes negative
following the same trends as seen in the fourth order cumulant.

A direct information on quark content of baryon carrying effective
degrees of freedom in the low temperature phase is also contained in
the $c_3/c_1$ ratio. At low temperature where thermodynamics is
dominated by three-quark modes $c_3/c_1=9$ for any value of
chemical potential. At low temperatures and at zero chemical potential $c_1$ and $c_3$ vanishes but
their ratio is finite. At asymptotically large temperatures  and for $\mu\to 0$
this ratio diverges. The ratio $c_3/c_1$ similarly as the kurtosis
$R_{4,2}=c_4/c_2$ exhibits strong variations in the phase transition
region.
It develops a peak with height which increases with $\mu/T$
and for sufficiently large $\mu/T$ it develops a deep structure. This
is the case for both mean-field and FRG calculations, however variations at
corresponding $\mu/T$ in the FRG are strongly suppressed because of  meson
fluctuations.



\subsection{Scaling properties of generalized susceptibilities at
  finite chemical potential}

The general trends and behavior of different susceptibilities
calculated in the last section can be understood considering their
scaling properties near the chiral transition. Under the mean-field  approximation
such scaling can be inferred from the Landau theory of phase
transitions, where the singular part of the thermodynamic potential is a
polynomial in an order parameter $\sigma$,
\begin{eqnarray}
  \nonumber
  \Omega^{sing}(T,\mu;\sigma) &=& \frac12 a(T,\mu) \sigma^2 + \frac14
  b(T,\mu) \sigma^4\\  &+& \frac16 c \sigma^6 -  h \sigma.
  \label{LG}
\end{eqnarray}
with $h$ being a symmetry breaking term.

%

For the chemical potential much smaller than the critical temperature
$T_c(\mu=0)$ of the second-order chiral phase transition the mass term
$a(T,\mu)$ is parameterized as
\begin{equation}
  a(T,\mu) = A\cdot(T-T_c)  + B_2\mu^2,
  \label{a}
\end{equation}
where both coefficients $A$ and $B_2$ are positive. In general, the
effective quartic coupling $b>0$, is also $T$ and $\mu$-dependent.
However, this dependence is irrelevant near the critical
line $a(T,\mu)=0$ and away from the CEP or the tri-critical point
(TCP).  Therefore in this case 
 we can also neglect the
higher order polynomial contribution by setting $c=0$.  In the chiral
limit $h=0$, the critical properties of $c_2$ and $c_4$ are
obtained from Eq.~(\ref{LG}):
\begin{eqnarray}
  c_2^{sing} &=& \frac{A B_2}{b T^2} \left( T - T_c\right) \theta( T_c -T  ), \\
  c_4^{sing} &=& \frac{6 B_2^2 }{b} \theta( T_c -T ).
  \label{c2c4LG}
\end{eqnarray}
Thus, $c_2$ is not differentiable at the critical temperature, while
$c_4$ has a discontinuity.

For a finite quark mass i.e. for $h\neq0$, the second-order
transition is turned to the crossover and the sharp structures in
$c_2$ and $c_4$ are smoothened. Therefore in the PQM model, the peak
structure appearing in $c_4$ is directly linked to the quark mass
and closely connected to the partial restoration of the chiral
symmetry.
From Eq.~(\ref{c2c4LG}) it is clear, that the kurtosis $R_{4,2}$,
driven by the $c_4$ has a discontinuity at $T_c$.

In the FRG approach the critical behavior of generalized
susceptibilities obtained under the mean-field approximation will be
modified by the long wave meson fluctuations. Detailed studies in the
QM model showed that the FRG method can correctly account for
long-range correlations resulting in the O(4) critical behavior of
  thermodynamic functions~\cite{SFR}. The gluon background is
not modifying critical dynamics related with the chiral symmetry.
Therefore in the chiral limit the PQM model belongs to the O(4)
universality class. The singular part of the thermodynamic
potential is controlled by the critical exponents of the
three-dimensional O(4)-symmetric spin system.  At vanishing chemical
potential
\begin{eqnarray}
  \Omega^{sing} &\sim& (T-T_c)^{2-\alpha},
\end{eqnarray}
leading to the following scaling of generalized susceptibilities
\begin{eqnarray}
  c_{2n}^{sing} &\sim& (T-T_c)^{2-n-\alpha}, \\
  c_{2n+1}&=&0
  \label{scaling}
\end{eqnarray}
for $n=1,2,3,\ldots$

In the mean-field approach the critical exponent $\alpha=0$.  The
quantum fluctuations within FRG renormalize the exponent to
$\alpha\simeq-0.21$ expected in the O(4) universality class. Therefore
fluctuations lead to weakening of  singularities. The finite quark mass
further smooths the temperature dependence of $c_2, c_3$ and $c_4$ as
seen in Figs.~\ref{fig:c2},~\ref{fig:c3} and~\ref{fig:c4}.  The
kurtosis follows singular behavior of $c_4$ and for $\mu=0$ exhibits a
step-like structure at $T_c$.


For finite chemical potential, but still away from the CEP or
the tricritical point (TCP), the coefficient $a(\mu,T)$ in the Landau
potential can be parameterized as
\begin{equation}
  a(T,\mu) = A\cdot(T-T_c)  + B_1\cdot(\mu-\mu_c)
  \label{a_fmu}
\end{equation}
while the quartic coupling $b>0$ and we still keep $c=0$.

In this case one gets the following expressions for  susceptibilities
\begin{eqnarray}
  \label{c2LG_fmu}
  c_1^{sing} &=& \frac{B_1 a }{2 T^3 b} \theta(-a), \\
  c_2^{sing} &=&  \frac{B_1^2}{2b T^2} \theta( -a  ), \\
  c_3^{sing} &=& c_4^{sing}= 0 .
  \label{c4LG_fmu}
\end{eqnarray}
Thus, $c_2$ exhibits a discontinuity while the singular part of $c_4$
is vanishing along with $c_3$ and higher order cumulants.  However,
there is the next to leading order contribution to $a(T,\mu)$ owing to
non-vanishing curvature of the transition line in the
$(T,\mu)$-plane. In order to take it into account one needs to add an
extra contribution $B_2\cdot(\mu-\mu_c)^2$ to the right-hand side of
Eq.~(\ref{a_fmu}).  In this case the $c_4^{sing}$ behaves as in
Eq.~(\ref{c2c4LG}) while the leading contribution to
$c_1^{sing}$ and $c_2^{sing}$ have the structure as in Eq.~(\ref{c2LG_fmu})
but with a modified $a(T,\mu)$. {The third order cumulant also develops
nonzero values in the broken phase $c_3^{sing}=\frac{3 B_2 B_1}{b T }\theta(-a)$.    }
As in the case of $\mu=0$, the kurtosis has a step-like behavior
but here the curvature of the phase diagram controls whether the
kurtosis is an increasing or decreasing function at the phase
transition because of two competing step structures in $c_2^{sing}$ and
$c_4^{sing}$.

Including quantum fluctuations as in the FRG approach the above MF
scaling is modified to the following O(4) relations
\begin{eqnarray}
  \Omega^{sing} &\sim& ( -a )^{2-\alpha} \\
  c_{n }^{sing} &\sim&  ( -a )^{2-n-\alpha}
  \label{scaling_fmu}
\end{eqnarray}
resulting in divergence of the $c_3$, $c_4$ and all higher order
cumulants along the O(4) critical line with the specific heat critical
exponent $\alpha\simeq -021$. The kurtosis is clearly divergent at
$T_c=T(\mu_c)$ following the singularity of the $c_4^{sing}$ cumulant.


The above scaling properties of the net-quark number fluctuations at
finite chemical potential are modified when approaching the TCP in the
chiral limit ar CEP at finite quark mass.

Close to the TCP the effective Landau potential has the structure as in
Eq.~(\ref{LG}) where the parameter $a(T,\mu)$ has a linear dependence on the
reduced temperature and chemical potential
\begin{equation}
  a(T,\mu) = A_a\cdot(T-T_c)  +  B_a\cdot (\mu-\mu_c).
  \label{aTCP}
\end{equation}
The quartic coupling tends to zero as
\begin{equation}
  b(T,\mu)=A_b\cdot (T-T_c) + B_b\cdot (\mu-\mu_c).
  \label{bTCP}
\end{equation}
The six-order coupling $c>0$, as it is required by the stability of
the theory.  Consequently, within Landaus theory and with the above
parametrization of the potential coefficients one gets the following
 relations for  the leading contribution to generalized susceptibilities,
\begin{eqnarray}
  c_1^{sing} &=& \frac{B_a}{2 T^3} \sqrt{\frac{-a}{c}} \theta(-a), \\
  c_n^{sing} &=& \frac{\Gamma(n-\frac32)}{2 \Gamma(\frac12)} \frac{(4c)^{n-2} B_a^n T^{n-4}}{(b^2-4ac)^{n-3/2}}, \ \ n>1.
  \label{c2c4LGTCP}
\end{eqnarray}
resulting in  the divergent kurtosis at the TCP as: $R_{4,2}^{sing} \sim
(b^2 - 4ac)^{-2}$.

From Eq.~(\ref{c2c4LGTCP}) one concludes that the $c_2^{sing}$ is
inversely proportional to a distance from the TCP along the
$a(T,\mu)=0$ line.  When approaching the TCP from any other direction
which is non-tangential to the critical line, then $b^2\ll a$ and
$c_2^{sing}$ is inversely proportional to the square root of the
distance to the TCP~\cite{CS,Hatta:2002sj}.  This demonstrates that the critical
region being defined by the properties of the $c_n$ for $n>1$ is
elongated along the O(4) critical line.

The thermodynamic properties and critical behavior near TCP are well
described by the mean-field theory up to logarithmic corrections,
because in this case the upper critical dimension is three.


For the non-zero external field (non-zero quark mass) the
three-critical point is turned to the CEP. Following the same
mean-field analysis (see also Ref.~\cite{Hatta:2002sj}) we obtain the
following leading singular behavior along the linear continuation of
the phase-coexistence line to the crossover region and near to the CEP
\begin{eqnarray}
  c_{n}^{sing}&\sim& |v|^{2-\frac32n}.
  \label{c2c4LGCEP1}
\end{eqnarray}
For any other directions which are asymptotically not equivalent to
the previous one
\begin{eqnarray}
 c_{n}^{sing}&\sim& |u|^{\frac43-n}.
  \label{c2c4LGCEP2}
\end{eqnarray}
where the introduced variables $v$ and $u$ are linear combinations of
$(T-T_{CEP})$ and $(\mu-\mu_{CEP})$.  For scaling (\ref{c2c4LGCEP1})
the kurtosis diverges as $R_{4,2}^{sing} \sim |v|^{-3}$, and for
scaling (\ref{c2c4LGCEP2}) the kurtosis $R_{4,2}^{sing} \sim
|u|^{-2}$.

When going beyond the mean-field approximation by including quantum
and thermal fluctuations in the PQM model one expects that along the
$u=0$ line the following scaling holds (see also~\cite{Stephanov:2008qz}):
\begin{eqnarray}
c_n^{sing} &\sim& |v|^{-[(n-2)(\gamma+\beta)+\gamma]},
  \label{c2c4LGCEP_sc_a}
\end{eqnarray}
and for any other direction
\begin{eqnarray}
c_n^{sing} &\sim& |u|^{-\left(n- 2 + \frac{\gamma}{\gamma+\beta} \right)}
  \label{c2c4LGCEP_sc_b}
\end{eqnarray}
Here, the critical exponents $\gamma$ and $\beta$ correspond to the
three-dimensional spin model belonging to the $Z(2)$ universality
class ~\footnote{In the Z(2) universality class the
  $\alpha\approx0.125$, $\beta=0.312$ and $\gamma=1.25$.}.  The
kurtosis is divergent at the CEP, however the strength of the
singularity depends on the direction. Approaching the TCP along the
$u=0$ line results in
\begin{eqnarray}
  R_{4,2}^{sing}&\sim& |v|^{  -2(\gamma+\beta) } =  |v|^{  -2 - \gamma+\alpha } ,
  \label{c2c4LGCEP_sc_R}
\end{eqnarray}
whereas for any other direction
\begin{eqnarray}
  R_{4,2}^{sing} &\sim&  |u|^{  -2  }.
  \label{c2c4LGCEP_sc_Ru}
\end{eqnarray}

The properties of the net-quark number density fluctuations and their
higher moments obtained in the PQM model at finite chemical potential
and the pion mass are qualitatively understood as remnants of the
critical structure and scaling behaviors related with the chiral
symmetry restoration in the limit of massless quarks.


\section{Summary and Conclusions}\label{sec:concl}
We have formulated thermodynamics of the Polyakov loop extended
quark--meson effective chiral model (PQM), including quantum
fluctuations within the functional renormalization group method
(FRG). We have solved the flow equation for the scale dependent
thermodynamic potential at finite temperature and density in the
presence of a background gluonic field.

We have shown that the non-perturbative dynamics introduced by the FRG
approach essentially modifies predictions of the model derived under
the mean-field approximation. In particular, we have demonstrated
quantitative changes of the phase diagram leading to a shift in the
position of the critical end point.

We have focused on fluctuations of the net-quark number density and
calculated the first four moments near the chiral transition for
different values of the chemical potential. We have indicated the role
and importance of ratios of different cumulant moments to identify
the deconfinement and chiral phase transitions.  We have also discussed
predictions of the Landau and scaling theories on the
critical behavior of the net-quark number density fluctuations and
their higher moments in the  vicinity of  the chiral phase
transition.

The extension of the FRG method proposed here to account for the
coupling of fermions to the background gluon fields within the
quark--meson model is of relevance to understand effectively
thermodynamics of the QCD near the chiral phase transition.


\section*{Acknowledgments}
V.~Skokov acknowledges the support by the Frankfurt Institute for
Advanced Studies (FIAS).  V.~Skokov thanks M.~Stephanov for valuable
discussions.  K. Redlich acknowledges partial support from the Polish
Ministry of Science.


\end{document}